\documentclass[10pt]{iopart}

\usepackage{iopams} 

\usepackage{graphicx}
\usepackage{bm}
\usepackage[latin1]{inputenc}
\usepackage{color}
\usepackage{hyperref}

\newcommand{\bra}[1]{\left\langle #1\right|}
\newcommand{\ket}[1]{\left| #1\right\rangle}

\newcommand{\Bra}{\langle}
\newcommand{\Ket}{\rangle}

\graphicspath{{./Figures/}}

\begin{document}

\review[Quantum Simulation with Interacting Photons]{Quantum Simulation with Interacting Photons}

\author{Michael J. Hartmann$^{1,2}$}

\address{$^{1}$Institute of Photonics and Quantum Sciences, Heriot-Watt University Edinburgh EH14 4AS, United Kingdom}
\address{$^{2}$Kavli Institute for Theoretical Physics, University of California, Santa Barbara, CA 93106, USA}
\ead{m.j.hartmann@hw.ac.uk}
\vspace{10pt}
\begin{indented}
\item[]April 2016
\end{indented}

\begin{abstract}
Enhancing optical nonlinearities so that they become appreciable on the single photon level and lead to nonclassical light fields has been a central objective in quantum optics for many years. After this has been achieved in individual micro-cavities representing an effectively zero-dimensional volume, this line of research has shifted its focus towards engineering devices where such strong optical nonlinearities simultaneously occur in extended volumes of multiple nodes of a network. Recent technological progress in several experimental platforms now opens the possibility to employ the systems of strongly interacting photons these give rise to as quantum simulators. Here we review the recent development and current status of this research direction for theory and experiment. Addressing both, optical photons interacting with atoms and microwave photons in networks of superconducting circuits, we focus on analogue quantum simulations in scenarios where effective photon-photon interactions exceed dissipative processes in the considered platforms.    
\end{abstract}

%
%
\submitto{\JOPT}
%
%
\ioptwocol
\tableofcontents

\section{Introduction}

Quantum simulation \cite{Buluta2009,Cirac12,Georgescu:2014ty} is a useful concept in several situations. One application considers the emulation of physical phenomena that are in their original form not accessible with existing experimental technology. This could for example be because the required energy or length scales exceed anything that is realisable. Another, possibly more prominent application of quantum simulations, which is the subject of this review is quantum many-body physics.  

The dimension of the Hilbert space for a quantum many-body system grows exponentially in the number of its constituents (the number of particles).
Hence, to specify an arbitrary quantum state of such a system one needs to store an exponentially growing amount of complex numbers, the probability amplitudes of the state's components. This task quickly becomes impossible on available classical computers. 
With numerical simulations of such systems being out of reach, one could hope to nonetheless explore them in detail via experiments.
Yet, unfortunately it is in many situations not possible to resolve their microscopic properties.
Here quantum simulation promises a way forward. Complex quantum many-body systems that defy experimental access are emulated by other systems which allow for much better experimental control and measurement resolution. The latter is often due to different temperature, energy, length or time scales at which quantum simulators work as compared to their simulation targets.
This idea to simulate a complex quantum system with another well controllable one was originally proposed by Feynman \cite{Feynman1982} by suggesting that a 'computer' built of quantum mechanical elements is needed to simulate highly complex quantum systems.

The observation that an exponentially growing amount of data is needed to specify a quantum state on the other hand indicates that quantum dynamics can execute computations with massive parallelism. Indeed there are known examples where a quantum computer could solve a problem exponentially faster than any known classical algorithm \cite{Ekert:1996yg}. 
In the mid 1990s, Lloyd showed that a universal quantum computer would also be able to simulate any quantum system efficiently \cite{Lloyd1996}.

With its conceptual use being well understood, the concept of quantum simulation received a boost in terms of practical implementations through the successful emulations of Bose-Hubbard models in equilibrium with ultra-cold atoms in optical lattices \cite{Greiner2002}. Quantum simulations with ultra-cold atoms have since then developed into a very active and successful field of research, see \cite{BDZ07} and \cite{Bloch:2012fu} for recent reviews.
Concepts for quantum simulation have also been successfully pursued with trapped ions \cite{Blatt:2012hb} and in quantum chemistry \cite{Kassal:2011uo,Lu:2012ek}. 

Quantum simulators can operate in two conceptually different ways. In digital quantum simulation, the time evolution to be simulated is decomposed into a sequence of short evolution steps using the Trotter formula. The individual steps of such a decomposition are often very similar to quantum gates, so that a digital quantum simulation can be viewed as a special purpose quantum computation. In analogue quantum simulation, in contrast, the device is operated in such a way that it's dynamics can, for suitable initial conditions, be approximated by an effective time-independent Hamiltonian, the physics of which one aims to simulate.

In this review we will summarise the developments towards quantum simulations with photons throughout recent years.
In doing so, we will concentrate on analogue simulations of quantum many-body physics with photons that are made to interact with each other via optical nonlinearities. We focus on regimes where the effective photon-photon interactions are a dominant process and in particular exceed the dissipation processes for individual photons.  We will thus only very briefly touch developments in digital quantum simulations but not cover the progress in simulations with linear optics devices and refer interested readers to the recent review by Aspuru-Guzik and Walter \cite{Aspuru-Guzik:2012eu}. 

Generating strong optical nonlinearities or effective photon-photon interactions has been a central goal of research in quantum optics for many years. Since more than a decade, optical nonlinearities on the single photon level have now been realised in single cavities that confine the trapped light fields as strongly as possible \cite{Walther:2006hc}. Due to technological progress in the micro-fabrication of high finesse resonators it has now become feasible to couple several resonators coherently to form an extended network \cite{Kimble}. Alternatively one-dimensional waveguides doped with optically highly nonlinear media can be considered. These developments give rise to networks or extended volumes where strong light-matter coupling and hence appreciable effective photon-photon interactions take place simultaneously in multiple locations.

The term ``interacting photons'' used in the title calls for some explanation. Pure photons do only exist in infinitely extended vacuum. Any matter that light enters into or any boundary conditions, e.g. in the form of a waveguide, it is subjected to will cause the emergence of a polarisation in these media. The elementary excitations of such fields are then polaritons, combinations of photons and excitations of the polarisation-fields, and these do interact if the medium has an appreciable non-linearity. It is photons in this sense that are considered in this review.   

The remainder of this review is organised as follows.
In section \ref{sec:optical-atoms}, we review the development towards quantum simulation with optical photons that are subject to strong effective interactions mediated by strongly polarisable atomic media or quantum well structures. We cover the research on cavity arrays leading to effective Bose-Hubbard or spin models and Jaynes-Cummings-Hubbard models, see section \ref{sec:optical-lattice}, as well as work on continuum models leading to Lieb-Lininger models and generalisations thereof, see section \ref{sec:optical-continuous}. Finally we briefly touch on approach employing atomic Rydberg media in section \ref{sec:Rydberg}. In section \ref{sec:microw-superc}, we then describe more recent developments with microwave photons hosted in networks of superconducting circuits. After briefly introducing the quantum theory for electrical circuits, c.f. section \ref{sec:microw-theory}, we cover developments towards Bose-Hubbard, c.f. section \ref{sec:microw-BH},  and  Jaynes-Cummings-Hubbard models, c.f. section \ref{sec:microw-JCH}, in these devices. We then point out some developments in digital quantum simulation, c.f. section \ref{sec:microw-DQS}, and discuss recent experimental progress, c.f. section \ref{sec:microw-exp}. In section \ref{sec:phase-diag}, we discuss the quantum many-body dynamics that can be explored in the considered quantum simulators. We first review equilibrium calculations of phase diagrams, c.f. section \ref{sec:phase-equi}, then discuss non-equilibrium and driven-dissipative regimes, c.f. section \ref{sec:non-equi}, and finally comment on employed and recently developed calculation techniques, c.f. section \ref{sec:calculation}, and the experimental signatures of the predicted phenomena, c.f. section \ref{sec:phase-diag-signatures}. We then discuss the more recent work on quantum simulation of many-body systems subject to artificial gauge fields, see section \ref{sec:artificial-gauge}, which has already seen some important experimental advances with photons and conclude with a summary in section \ref{sec:summary}.

\section{Optical photons interacting with atoms}
\label{sec:optical-atoms}

First theory developments towards quantum simulations with interacting photons considered optical photons that couple to atoms.
In order to generate sufficiently strong optical nonlinearities for this aim, strongly confined light-fields with appreciable interactions between individual photons and atoms where considered. Such strong atom-light interactions had been achieved in high finesse optical cavities \cite{Birnbaum05} in the years before and so multiple cavities coupled via tunnelling of photons between them were considered initially, see also the reviews by Hartmann et al. \cite{Hartmann08}, Tomadin et al. \cite{Tomadin2010} and Noh et al. \cite{Noh16}. Yet, for optical frequencies, the strength of achievable light matter couplings is about six orders of magnitude smaller than the frequency of the employed photons and it is thus quite demanding to build multiple cavities with sufficiently low disorder so that photons can tunnel efficiently between them. As a consequence setups with thinly tapered optical fibres have been considered subsequently to avoid these challenges, see also a recent review by Roy et al. \cite{Roy16}. We here first review the work on cavity arrays and then turn to discuss continuum models in tapered optical fibres.

\subsection{Lattice models in cavity arrays}
\label{sec:optical-lattice}

\subsubsection{Coupling of cavities}
\label{sec:cca}
\begin{figure}
	\centering
	\includegraphics[width=\linewidth]{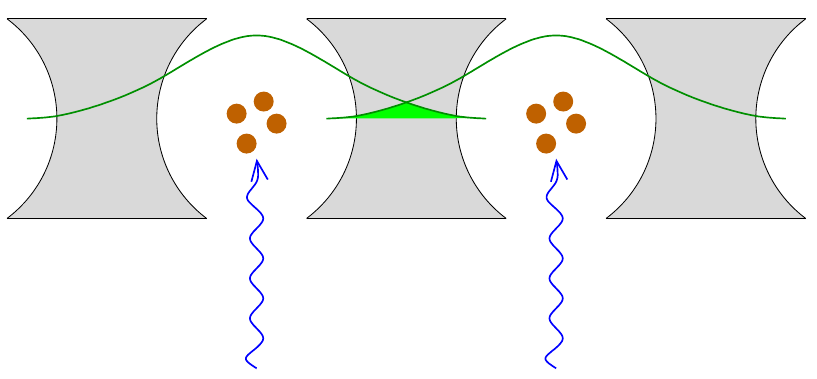}
	\caption{An array of nonlinear cavities. Photon hopping occurs due to the overlap (shaded green) of the light modes (green lines) of adjacent cavities. Atoms in each cavity (brown), which are driven by external lasers (blue) give rise to an on site potential. Reprinted from \cite{Hartmann06}.}
	\label{fig:crystal}
\end{figure}

In cavity arrays, the individual cavities are coupled via tunnelling of photons between them due to the overlap of the spacial profile of their resonance 
modes, c.f. figure \ref{fig:crystal}. Following reference \cite{Yariv99} an array of cavities can be described by a periodic dielectric constant, $\epsilon (\vec{r}) = \epsilon (\vec{r} + R\vec{n})$, where $\vec{r}$ is the three dimensional position vector, $R$ the lattice constant (i.e. the distance between the centres of adjacent cavities) and $\vec{n}$ a vector of integers. Expanding the electromagnetic field in terms of Wannier functions $w_{\vec{R}}$, localised in the cavities at locations $\vec{R} = R \vec{n}$, the tunnelling rate of photons between neighbouring cavities can be expressed as 
\begin{equation} \label{eq:bare-photon-hopping}
J_{phot} = 2 \omega_C \int d^3r \, \left[\epsilon_{\vec{R}}(\vec{r}) \, - \, \epsilon(\vec{r}) \right] \,
\vec{w}_{\vec{R}}^{\star} \vec{w}_{\vec{R}'},
\end{equation}
where $|\vec{R} - \vec{R}'| = R$, $\omega_{C}$ is the resonance frequency of the considered cavity mode and the dielectric function $\epsilon_{\vec{R}}(\vec{r})$ describes a single cavity surrounded by bulk material only.

Introducing creation and annihilation operators $a_{\vec{R}}^{\dagger}$ and $a_{\vec{R}}$ for the Wannier modes, the Hamiltonian of the field can thus be written as
\begin{equation} \label{eq:arrayham2}
\mathcal{H} = \omega_C \sum_{\vec{R}}
a_{\vec{R}}^{\dagger} a_{\vec{R}} + J_{phot} \sum_{<\vec{R}, \vec{R}'>} a_{\vec{R}}^{\dagger} a_{\vec{R}'},
\end{equation}
where $\sum_{< \vec{R}, \vec{R}' >}$ is a sum over nearest neighbours. Since the tunnelling rate is typically much less than the photon frequency, $J_{phot} \ll \omega_{C}$, a rotating wave approximation has been applied. 
Equation (\ref{eq:arrayham2}) assumes that all the cavities have the same resonance frequency and that the tunnelling rate is the same for all cavity-cavity interactions. In practise there will always be some disorder in the array
and the resonance frequencies, $\omega_C (\vec{R})$, and tunnelling rates, $J_{phot}(\vec{R},\vec{R}')$ will differ from cavity to cavity. This disorder in the array is a significant challenge for experimental realisations, see section \ref{sec:exp-req} for further discussion. On the other hand it can even give rise to interesting effects such as the emergence of glassy phases, see section \ref{sec:phase-diag}.


\subsubsection{Bose-Hubbard Models}
\label{sec:optical-BoseHubbard}

Although the Hamiltonian (\ref{eq:arrayham2}) has applications for guiding light-modes in ``coupled resonator optical waveguides'', see \cite{Yariv99}, it is of limited interest in the context of quantum simulation.
Indeed, the model is fully harmonic and can thus be solved efficiently by for example deriving equations of motion for the first and second order moments of the creation and annihilation operators in the Heisenberg picture. Such non-interacting models thus do not feature a significant quantum complexity and no quantum simulators are required to obtain understanding of them. Nonetheless several interaction free models have recently received significant interest for experiments as they can model artificial gauge fields and give rise to peculiar band structures and chiral edge modes. We discuss such models in more detail in section \ref{sec:artificial-gauge}.

The complexity of a model in turn grows dramatically if interactions between excitations play a role and usually no exact solutions are known in such cases.
Here we in particular consider on-site interactions as their interplay with the tunnelling leads to interesting many-body physics. A paradigm for such a situation is the Bose-Hubbard Hamiltonian,
\begin{eqnarray} \label{eq:bosehubbard}
H_{BH} = \mu \sum_{\vec{R}} p^{\dag}_{\vec{R}} p_{\vec{R}} &-& J \sum_{<\vec{R}, \vec{R}'>} (p_{\vec{R}}^{\dag}p_{\vec{R}'} + \textrm{H.c.}) \\ 
&+& \frac{U}{2} \sum_{\vec{R}} p^{\dag}_{\vec{R}} p_{\vec{R}} (p^{\dag}_{\vec{R}} p_{\vec{R}} - 1),\nonumber
\end{eqnarray}
where $p_{\vec{R}}^{\dag}$ creates a boson at site $\vec{R}$, $J$ is the hopping rate, $U$ the on-site interaction strength, and $\mu$ the chemical potential. Motivated by its application to Josephson junction arrays \cite{Fazio2001}, this model received significant interest in the 1990s and its ground state phase diagram has been discussed by Fisher et al. \cite{Fisher89}. It is
characterised by two different phases at zero temperature, an incompressible Mott insulating phase in the interaction dominated regime, $U > J$, with commensurate filling and a superfluid phase elsewhere. The phase transition between both phases has been observed in a seminal experiment with ultra-cold atoms in an optical lattice \cite{Greiner2002,BDZ07,Bloch:2012fu}.

Approaches to generate an effective Bose-Hubbard model for quantum simulation purposes with photonic excitations consider polaritons that are formed by photons which either interact with atoms \cite{Hartmann06} or with quantum well excitons \cite{Verger:2006kq}. We first discuss setups involving atom-photon interactions.

\subsubsection{Bose-Hubbard model with dark state polaritons}
\label{sec:pbh}
\begin{figure}
	\centering
	\includegraphics[width=.8\linewidth]{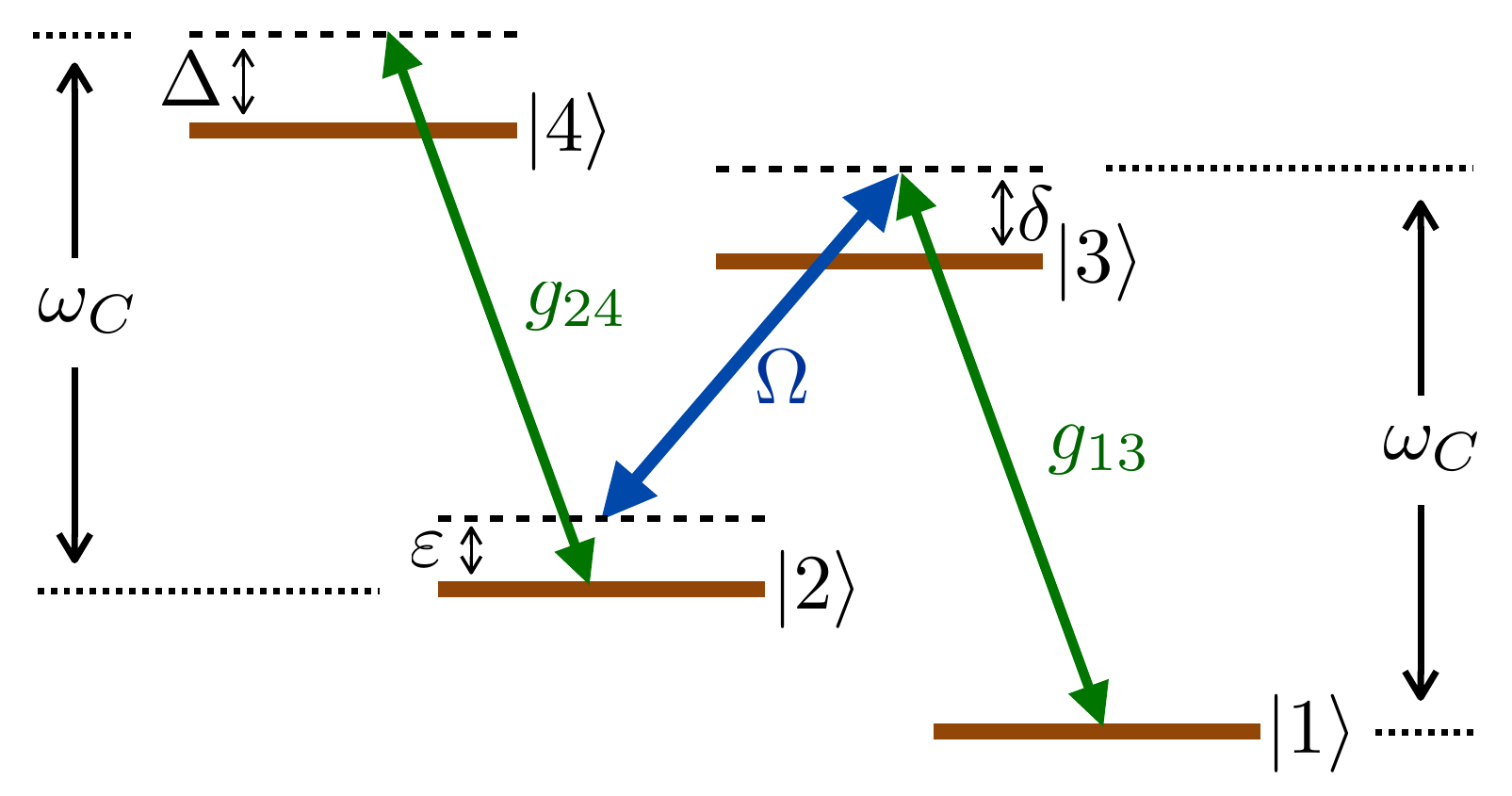}
	\caption{The level structure and the possible transitions of one atom, $\omega_C$ is the frequency of the cavity mode, $\Omega$ is the Rabi frequency of the driving by the laser, $g_{13}$ and $g_{24}$ are the parameters of the respective dipole couplings and $\delta$, $\Delta$ and $\varepsilon$ are detunings.}
	\label{fig:level}
\end{figure}

A Bose-Hubbard model can be generated in an array of cavities that are filled with atoms of a specific four-level structure \cite{Hartmann06} and driven by a laser in the same manner as in Electromagnetically Induced Transparency (EIT) \cite{Fleischhauer05}, see figure \ref{fig:level}. The transitions between levels $\ket{2}$ and $\ket{3}$ couple to the laser field whereas the transitions $\ket{2} \leftrightarrow \ket{4}$ and $\ket{1} \leftrightarrow \ket{3}$ couple to the cavity resonance mode. Levels $\ket{1}$ and $\ket{2}$  are assumed to be metastable with negligible spontaneous emission rates.

As has been shown by Imam\u{o}glu and co-workers, this atom cavity system can exhibit a very large optical nonlinearity \cite{Imamoglu97,Werner99} leading to the phenomenon of photon blockade, where an input drive that is resonant to the first excitation in the system can only generate one excitation which needs to decay before a subsequent excitation can be generated \cite{Imamoglu97,Werner99}. Photon blockade has so far been experimentally shown with coherent \cite{Birnbaum05,Faraon:2008cr,Reinhard:2012xe,Englund:2012jo,Bozyigit:2011fx,Lang2011} as well as incoherent driving \cite{Hoffman2011}.

For one cavity filled with $N$ atoms of the level structure sketched in figure \ref{fig:level}
the Hamiltonian describing the atom-photon interactions reads
\begin{eqnarray} \label{H_manyatom}
H^I & = &
\sum_{j=1}^N \left(\varepsilon \sigma_j^{22} + \delta \sigma_j^{33} + (\Delta + \varepsilon)
\sigma_j^{44} \right) \\
& + & \, \sum_{j=1}^N \left( \Omega \, \sigma_j^{23} \, + \,
g_{13} \, \sigma_j^{13} \, a^{\dagger} \, + \,
g_{24} \, \sigma_j^{24} \, a^{\dagger} \, + \, \textrm{H.c.} \right)  \, , \nonumber
\end{eqnarray}
in a rotating frame with respect to
$H_0 = \omega_C \left( a^{\dagger} a + \frac{1}{2} \right) + \sum_{j=1}^N \left( \omega_C \sigma_j^{22} + \omega_C \sigma_j^{33} + 2 \omega_C \sigma_j^{44} \right)$.
Here $\sigma_j^{kl} = \ket{k_j}\bra{l_j}$ is the transition operator between levels $\ket{l}$ and $\ket{k}$ of atom $j$, $\omega_C$ is the frequency of the cavity mode, $\Omega$ is the Rabi frequency of the laser drive and $g_{13}$ ($g_{24}$) are the coupling strengths of the cavity mode to the atomic transitions  $\ket{2} \leftrightarrow \ket{4}$ ($\ket{1} \leftrightarrow \ket{3}$). 

A cavity array as described in section \ref{sec:cca}, where each cavity is doped with $N$ four-level atoms as depicted in figure \ref{fig:level}, can form a quantum simulator for a Bose-Hubbard model, c.f. equation (\ref{eq:bosehubbard}), if all atoms interact in the same way with the cavity mode and the number of excitations is significantly lower than the number of atoms in each cavity. In this regime, the dynamics generated by the Hamiltonian (\ref{H_manyatom}) can be described in terms of polaritons, hybridised light-matter quasi-particles that obey bosonic statistics. 
One species of these polaritons does only occupy the atomic levels which do not have a direct dipole transition to the atomic ground state. These polaritons have been considered by Fleischhauer et al.  \cite{Fleischhauer05} and are called dark state polaritons as they do not lead to emission of radiation and are therefore long lived. The latter property is obviously rather beneficial for quantum simulation applications. The creation operators of the dark state polaritons read
\begin{equation} \label{polariton_operators}
p^{\dagger} = \frac{1}{B} \, \left(g S_{12}^{\dagger} - \Omega a^{\dagger} \right)
\end{equation}
where $g = \sqrt{N} g_{13}$ is a collective coupling rate, $S_{12}^{\dagger} = \frac{1}{\sqrt{N}} \sum_{j=1}^N \sigma_j^{21}$ creates a spin wave in the metastable atomic levels and $B = \sqrt{g^2 + \Omega^2}$.
The dynamics of the dark state polaritons decouples from
the remaining species for a suitable parameter regime, where the frequencies of the species are sufficiently separated.

The coupling of the dark state polaritons to level $\ket{4}$ of the atoms induces an effective interaction between all dark state polaritons in one cavity. For $|g_{24 }g \Omega / B^2| \ll |\Delta|$ the strength of this interaction can be calculated in a perturbative manner \cite{Werner99},
\begin{equation} \label{eq:osint}
U_{BH} = \frac{2 g_{24}^2}{\Delta} \,
\frac{N g_{13}^2 \, \Omega^2}{\left(N g_{13}^2 \, + \, \Omega^2 \right)^2}.
\end{equation}
Similarly, the two photon detuning $\varepsilon$ leads to an energy shift of $\mu_{BH} = \epsilon g^2 / B^2$ for the polaritons that plays a similar role as a chemical potential, see also section \ref{sec:microw-chem-pot}. 

Provided the hopping rate of photons between cavities is small compared to the frequency separation between the polariton species, the hopping of photons translates into a hopping of dark state polaritons at a rate \cite{Hartmann06}
\begin{equation} \label{eq:Jvalue}
J_{BH} = \frac{2 \Omega^2}{N g_{13}^2 + \Omega^2} J_{phot} \, 
\end{equation}
where $J_{phot}$ is defined in equation (\ref{eq:bare-photon-hopping}).

As a consequence the Hamiltonian for the dark state polaritons takes on the form of a Bose-Hubbard model as introduced in equation (\ref{eq:bosehubbard}), with $J=J_{BH}$ [c.f. equation (\ref{eq:Jvalue})], $U=U_{BH}$ [c.f. equation (\ref{eq:osint})] and $\mu = \mu_{BH}$. Note that both,  $J_{BH}$ and $U_{BH}$ can be tuned by varying the intensity of the laser driving $\Omega^2$. See also \cite{Hartmann:2008ef} for a generalisation to two polariton species and \cite{Brandao:2008zm,Nikoghosyan:2012aq,Ridolfo:2013yb} for alternative nonlinearities.

Concepts for quantum simulators of Bose-Hubbard models have also been put forward for a rather different experimental platform based on semiconductor structures. We discuss these in the next section.

\subsubsection{Bose-Hubbard model with exciton polaritons}
\label{sec:exciton-pol}
An interaction of the form as in the Bose-Hubbard model, see equation (\ref{eq:bosehubbard}), was also shown to exists and analysed for polaritons formed by photons and excitons of a quantum well \cite{Verger:2006kq}. The Hamiltonian describing the interactions between the photons of one cavity mode and the excitons of the matching wave-vector here reads,
\begin{equation} \label{eq:exciton-photon-int}
H = \omega_X b^\dag b + \frac{U_X}{2} b^\dag b^\dag b b + \omega_C a^\dag a + g (a b^\dag + a^\dag b ),
\end{equation}
where $a^\dag (a)$ creates (annihilates) a cavity photon and $b^\dag (b)$ creates (annihilates) an exciton.
The excitons have transition frequencies $\omega_X$, interact with each other at a strength $U_X$ and couple to photons at a strength $g$.
An anharmonic exciton-photon coupling that depends on the exciton oscillator strength saturation density can typically be neglected compared to the interaction between excitons \cite{Verger:2006kq}. 

In a so called ``strong coupling regime`` where $g$ is larger than the linewidth of the cavity and the excitons, polaritons, i.e. hybridised light-matter excitations, become the elementary excitations of the system. These have annihilation operators
\begin{eqnarray}\label{eq:PolaritonModesBH}
p_+&=&\cos(\theta)a+\sin(\theta)b\\
p_-&=&\sin(\theta)a-\cos(\theta)b\nonumber,
\end{eqnarray}
and frequencies
\begin{equation} \label{eq:PolaritonFrequsBH}
\omega_{\pm} = \frac{\omega_X + \omega_C}{2} \pm \frac{\sqrt{(\omega_C - \omega_X)^2 + g^2}}{2}
\end{equation}
with $\sin(\theta) = g/\sqrt{g^2 + \tilde{\Delta}^2}$,
$\cos(\theta) = \tilde{\Delta}/\sqrt{g^2 + \tilde{\Delta}^2}$ and $\tilde{\Delta} = \omega_C - \omega_X +\sqrt{(\omega_C - \omega_X)^2 + g^2}$.

Since $g \gg U_X$ the dynamics of both polariton species decouples from each other and the interaction between excitons leads to an interaction $ (U_{p-}/2) p_-^\dag p_-^\dag p_- p_-$ with $U_{p-} = U_X \cos^4(\theta)$ for the lower polaritons $p_-$.   
These interactions can have a strength comparable or even larger than their linewidth due to the strong interactions of the excitonic components of the polaritons. Ways to enhance these polariton-polariton interactions via Feshbach type resonances between polariton pairs and bi-excitons have been explored by Carusotto et al. \cite{Carusotto:2010hb}.

By confining the light modes in a periodic structure, a lattice model as given in equation (\ref{eq:bosehubbard}) can be generated \cite{Byrnes:2010kb}. The periodic confinement for the photons can thereby be either a photonic crystal or a structure of connected micro-pillars.
Such polariton-polariton interactions and their interplay with polariton tunnelling between lattice sites has recently been investigated in a self trapping experiment \cite{Abbarchi:2013ix}, see section \ref{sec:self-trap} for a more detailed discussion.

Besides Bose-Hubbard physics, most of the research effort on lattice models with optical photons has considered scenarios where the photons couple to one two-level system in each cavity. We review these approaches in the next section.  

\subsubsection{Jaynes-Cummings-Hubbard model}
\label{sec:optical-JCHubbard}
An effective repulsive interaction between photons or polaritons can also be generated by doping a cavity with one atom of which only one internal transition couples to the cavity resonance mode. This situation of a two-level emitter coupled to a single cavity mode is described by the celebrated Jaynes-Cummings model \cite{Jaynes1963},
\begin{equation} \label{JC_JCH}
H^{JC} = \omega_C a^{\dag}a + \omega_0 \ket{e}\bra{e} + g (a^{\dag}\ket{g}\bra{e} + a\ket{e}\bra{g}),
\end{equation}
where $\omega_C$ and $\omega_0$ are the frequencies of the cavity mode and the atomic transition, $g$ is Jaynes-Cummings light-matter coupling, $a^{\dag}$ is the creation operator of a photon in the resonant cavity mode, and $\ket{g} (\ket{e})$ are the ground (excited) states of the two level system.
Since the energy of a photon $\omega_C$ and the atomic transition energy $\omega_0$ are much greater than the coupling $g$, the number of excitations is conserved by the Hamiltonian (\ref{JC_JCH}). Hence it can be diagonalised for each manifold with a fixed number of excitations $n$ separately. The energy eigenvalues $E_n$ for $n$ excitations read $E_0 = 0$ and
$E_n^{\pm} =  n \omega_C + \frac{\Delta}{2} \pm \sqrt{n g^2 + \frac{\Delta^2}{4}}$ 
for $n \ge 1$, where $\Delta = \omega_0 - \omega_C$. As a consequence, the energy of the lowest state with two excitations is not twice the energy of a single excitation state and their difference 
\begin{equation} \label{effJCrep}
U_{JC} = E_2^- - 2 E_1^- = 2 \sqrt{g^2 + \frac{\Delta^2}{4}} - \sqrt{2 g^2 + \frac{\Delta^2}{4}} - \frac{\Delta}{2}
\end{equation}
plays the role of an effective on-site repulsion that can be tuned via the detuning $\Delta$.
Photon blockade due to the effective repulsion (\ref{effJCrep}) has been observed for an individual atom \cite{Birnbaum05}, quantum dots in photonic crystals \cite{Faraon:2008cr,Reinhard:2012xe,Englund:2012jo} and a circuit quantum electrodynamics (circuit QED) setup \cite{Bozyigit:2011fx,Lang2011}, see also section \ref{sec:microw-superc} for further details. 

The concept of photon blockade in a Jaynes Cummings model can also be explored for higher excitation numbers, where the difference between the energy of $n+1$ excitations, $E_{n+1}^{-}$, and the energies of $n$ excitations, $E_{n}^{-}$, and a single excitation, $E_{1}^{-}$, should be considered. For $\Delta = 0$ one finds
\begin{equation} \label{effJCrep-n}
U_{JC;n} = E_{n+1}^{-} - (E_{n}^{-} + E_{1}^{-}) = g \, (\sqrt{n} + 1 - \sqrt{n+1}), 
\end{equation}
so that the interaction energy per excitation decreases with growing excitation number $n$, $U_{JC;n}/n \to g/n$ for $n \gg 1$.
The degrading of the anharmonicity of the spectrum of the Jaynes-Cummings model at high excitation numbers was observed in a self-trapping experiment \cite{Raftery:2014fj}, see Sec. \ref{sec:self-trap}, and the resulting breakdown of the photon blockade effect was investigated by Carmichael \cite{Carmichael:2015qf} as an example for a dissipative quantum phase transition. He found the transition to be first order, as indicated by a bi-modality of the Q-function, except for a critical
value of drive strength for zero drive detuning, where it is a continuous second order transition.

In an array of cavities where each cavity is doped with a single two-level emitter that is coupled to the light mode as described by equation (\ref{JC_JCH}),
the effective repulsion of equation (\ref{effJCrep}) will suppress the mobility of excitations in a similar manner as the on-site interactions in the Bose-Hubbard model since the lowest energy for two excitations in one cavity is $E_2^-$ and moving one additional excitation to a cavity requires an extra energy of $U_{JC}$.
This setup has first been investigated by Angelakis et al. \cite{Angelakis2007} and Greentree et al. \cite{Greentree2006}.
The full effective many-body model in a cavity array that is based on the effective interaction
(\ref{effJCrep}) has been coined {\it Jaynes-Cummings-Hubbard} model and reads,
\begin{equation} \label{eq:JCHubbard}
H_{JCH} = \sum_{\vec{R}} H^{JC}_{\vec{R}}
- J_{JC} \sum_{<\vec{R}, \vec{R}'>} \left( a_{\vec{R}}^{\dagger} a_{\vec{R}'} + \textrm{H.c.} \right),
\end{equation}
where $\vec{R}$ labels the site of a cavity. Each cavity contains one atom interacting via the Jaynes-Cummings interaction with the cavity mode and photons tunnel between neighbouring cavities at a rate $J_{JC} = J_{phot}$. The dispersive regime of the Jaynes-Cummings-Hubbard and Rabi-Hubbard model was investigated in \cite{Zhu:2013wd}.

\paragraph{Multiple two-level atoms per cavity}
Effective many-body physics based on the Hamiltonian (\ref{eq:JCHubbard}) can not only be observed for a single two level system in each cavity, as assumed in equation (\ref{JC_JCH}), but also for setups with several two level systems per cavity. Such a model can describe photonic crystal micro-cavities doped with substitutional donor or acceptor impurities. This approach to implementing effective many-body models, which can have suitable parameters, has been proposed in \cite{Na2008}. The phase transitions of a model with several two level systems in each cavity have also been studied in \cite{Rossini2007,Lei2008}, see section \ref{sec:phase-diag} for further discussions.

\subsubsection{Spin Models}
Although it is not the main topic of this review, we note here that coupled cavity arrays have also been considered for the simulation of spin lattice Hamiltonians  \cite{Angelakis2007,Hartmann:2007qc,Hartmann08,Gonzalez-TudelaA.:2015wu,Aron:2016zf}. Here the internal levels of the atoms in the cavities represent the spin degrees of freedom and interactions between spins can be mediated via off-resonant couplings to collective photon modes \cite{Hartmann:2007qc,Hartmann08,Zhu:2013wd,Gonzalez-TudelaA.:2015wu,Aron:2016zf}.

After having discussed the main strands of the theory work on simulating quantum lattice Hamiltonians with optical photons, we now take a closer look at the requirements and challenges for implementing sthese approaches in experiments.

\subsubsection{Experimental requirements for lattice models}
\label{sec:exp-req}
Electromagnetic excitations, including polaritons and photons, inevitably couple to the electromagnetic vacuum that can not be excluded from any experiment. These excitations will therefore always by limited to a finite lifetime or trapping-time.
In order to be able to explore their dynamics, they need to be kept in the experimental sample for a time that exceeds the timescales associated to the kinetic and interaction energies. Denoting the rate of photon losses from the cavities by $\kappa$ and the rate of spontaneous emission for the atoms by $\gamma$, one needs
\begin{equation} \label{eq:targetregimelattice}
U_{BH} , J_{BH} > \kappa, \gamma  \quad \textrm{or} \quad U_{JC}, J_{JC} > \kappa, \gamma 
\end{equation}
for effective Bose-Hubbard or Jaynes-Cummings-Hubbard models. In terms of the light matter couplings, meeting the conditions (\ref{eq:targetregimelattice}) requires both transitions to operate at high cooperativity for single photons, $g_{13}^2 \gg \kappa \gamma$ and $g_{24}^2 \gg \kappa \gamma$, for the Bose-Hubbard model, or a strong coupling regime, $ g \gg \kappa, \gamma$ for the Jaynes-Cummings-Hubbard model. For the approach to realise a Bose-Hubbard model as explained in section \ref{sec:optical-BoseHubbard} the condition $\Delta > \gamma$ is also needed. The fact that for this model only the product of the two decay rates $\kappa$ and $\gamma$ needs to be bounded can be understood by realising that the relative contributions of the photonic and atomic components in the dark state polaritons, c.f. equation (\ref{polariton_operators}) can be varied to avoid the faster decay channel and optimise their lifetime. 

An alternative approach to the photon blockade effect was discovered in a two-site Bose-Hubbard system where a resonantly driven nonlinear cavity is tunnel-coupled to an auxiliary cavity. Remarkably this leads to anti-bunched output photons even if $U\ll \kappa$ \cite{Liew2010} but requires low rates of pure dephasing $\gamma^{*}$ \cite{Ferretti:2013fk}. The effect is due to a destructive interference of two excitation paths as can be seen from an expansion in excitation numbers at low drive intensities \cite{Bamba2011}. The direct path to generate a second excitation in cavity one via the coherent input, $|1,0\rangle \to |2,0\rangle$, destructively interferes with the excitation path where the first excitation tunnels to the second cavity, $|1,0\rangle \to |0,1\rangle$, a second excitation is generated in cavity one via the drive, $|0,1\rangle \to |1,1\rangle$, and the excitation in cavity two tunnels back to cavity one. Its origin in an interference of excitation paths explains why this photon blockade effect requires that the nonlinearity exceeds the rate of pure dephasing $U \gg \gamma^{*}$. Despite these requirements for an experimentally suitable technology, there has already been significant progress as we discuss next.

\subsubsection{Experimental progress}
\label{sec:self-trap}
Demonstrating coherent coupling between high finesse optical resonators with low enough disorder and light-matter interactions in the strong coupling regime at the same time remains an experimental challenge at the time of writing. Yet, coupled arrays of photonic band-gap cavities in photonic crystal structures that are suitable for quantum simulation applications have been built with 10-20 cavities \cite{majumdar12c,Jarlov:2013lq}. A further possibility to engineer a tunnel coupling between adjacent cavities is to connect these via waveguides that come close to each other at a coupling point \cite{Lepert:2011kz}. In this way a controllable coupling can be combine with a small mode volume but open cavity that allows to strongly couple laser trapped atoms to its resonance modes.  

\paragraph{Realization of the Jaynes-Cummings-Hubbard model with trapped ions}
A minimal version of the Jaynes-Cummings-Hubbard model with two lattice sites has been realised with trapped ions \cite{Toyoda:2013jl}. Here the motion of the ions was employed to implement the harmonic degree of freedom thus replacing the cavity modes and the internal levels of the ions represent the two-level systems. Hence the creation and annihilation operators $a_j^\dag$ and $a_j$ here describe phonons rather than photons. In the experiment, an adiabatic sweep from a regime dominated by the on-site repulsion $U_{JC}$ to a regime dominated by the tunnelling $J_{JC}$ was performed via varying the detuning $\Delta$ by tuning the frequency of a laser that drives a red side-band transition for the ions. The mobility of the polaritons in the tunnelling dominated regime and the suppression of their mobility in the interaction dominated regime was then shown by measuring their on-site number fluctuations.

\paragraph{Self-trapping experiments}
The interplay of interactions and tunnelling between adjacent lattice sites has also been investigated in two experiments with two coupled resonators \cite{Abbarchi:2013ix,Raftery:2014fj}. These experiments explored self-trapping effects in regimes of high excitation densities, an interaction phenomenon that already becomes accessible for moderate interactions between individual excitations. The effect appears for two coupled resonators with a strong imbalance of excitation numbers, where the interaction energies per excitation differ between both resonators by an amount that exceeds the rate of inter-resonator tunnelling. 
\begin{figure*}[h]
\includegraphics[width=\textwidth]{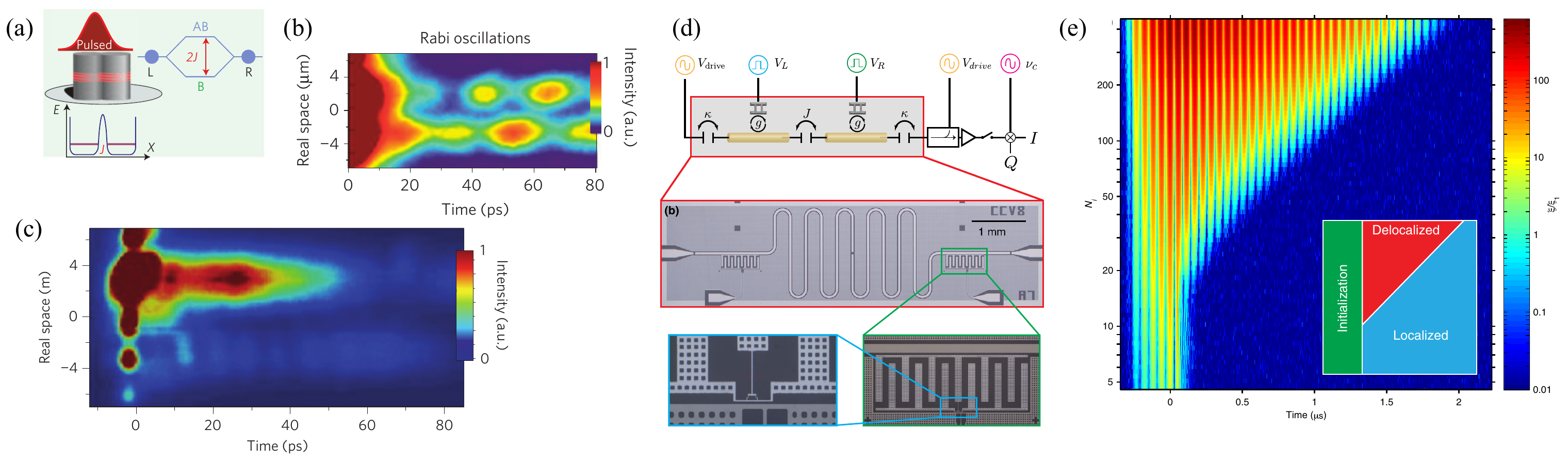}
\caption{Self-trapping experiments: (a) Set-up of two coupled micro-pillars containing Bragg stack cavities that couple via an overlap of their trapped photon modes \cite{Abbarchi:2013ix}. Polaritons are initially generated in the left pillar. (b) Collective oscillations of the polaritons between both pillars as measured by the emitted intensities for moderate initial polariton imbalance. (c) Self trapped regime where the initial polariton imbalance is strong enough to suppress collective oscillations. (d) Set-up of two coupled superconducting resonators that each interact with a transmon qubit \cite{Raftery:2014fj}. (e) Phase diagram as observed in the experiment \cite{Raftery:2014fj} with collective oscillations for large initial imbalance that are suppressed as the imbalance decreases. Plots a,b and c adapted from \cite{Abbarchi:2013ix} and plots d and e adapted from \cite{Raftery:2014fj}}
\label{fig:selftrapping}
\end{figure*}

For a nonlinearity of the form $U n(n-1)$ [$n$ is the number of particles], which has been realised in an experiment with exciton polaritons in two coupled Bragg stack micro-pillars \cite{Abbarchi:2013ix}, the interaction energy per particle is $U(n-1)$ and self trapping occurs for high particle densities but ceases as the particle number decays. In their experiment, Abbrachi et al. \cite{Abbarchi:2013ix} thus observed a transition from a self-trapped regime to a regime of excitation oscillations between the resontors as the particle number decreased over time due to dissipation, see figure \ref{fig:selftrapping}a-c. 

For interactions as present in the Jaynes-Cummings model, the interaction energy per excitation degrades as the number of excitations grows, see Eq. (\ref{effJCrep-n}). One thus observes oscillations of excitations between both resonators for a strong initial excitation number imbalance \cite{Schmidt2010a}. This has been seen by Raftery et al. in an experiment with two coupled superconducting coplanar waveguide resonators that each interact with a transmon qubit \cite{Raftery:2014fj}, c.f. section \ref{sec:microw-superc}. As the excitation number decreased a transition to the self-trapped regime was observed, see figure \ref{fig:selftrapping}d-e. 

After reviewing work on the simulation of quantum lattice models, we now turn to discuss efforts towards quantum simulators for continuum models. Importantly, for photons at optical frequencies these approaches face less experimental challenges.

\subsection{Continuum Models in Optical Fibers}
\label{sec:optical-continuous}

For optical frequencies, building mutually resonant cavities of sufficient finesse is very challenging. This can be appreciated by observing that the largest achievable atom-photon couplings reach 1 - 10 GHz \cite{Walther:2006hc}. Hence disorder in the resonance frequencies of the cavities needs to be suppressed to $10^9$ Hz or below, which corresponds to a disorder in cavity dimensions below $10^{-6}$ times the wavelength of the trapped photons. A possible alternative to cavity arrays are therefore one-dimensional waveguides, which avoid the need to build mutually resonant cavities but nonetheless feature a large light-matter coupling due to a tight confinement of the light modes in transverse directions. Moreover, in contrast to lattice structures, such devices emulate a different class of quantum many-body models. The probably most prominent representative of this class is the Lieb-Liniger model,
\begin{equation}\label{eq:LLmodel}
H_{LL} = \int_0^L dz \left[\frac{\hbar}{2 m_{\textrm{eff}}} (\partial_z \psi^\dag) (\partial_z \psi)  + \frac{\tilde{g}}{2} (\psi^\dag)^2 \psi^2 \right]
\end{equation}
which describes bosons of effective mass $m_{\textrm{eff}}$ in one dimension which interact via a contact interaction of strength $\tilde{g}$ ($\hbar$ is Planck's constant). In equation (\ref{eq:LLmodel}), $L$ is the length of the waveguide and $\psi$ the field describing the polaritons.

All ground state features of the Lieb-Liniger model \cite{Lieb:1963fq}   
are characterised by a single, dimensionless parameter 
$G = m_{\textrm{eff}} \tilde{g}/(\hbar^2 N_p/L)$ [$N_p$ 
is the particle number], that quantifies the 
effective interaction strength between the particles. For weak interactions, i.e. $|G| < 1$, the bosons are in a superfluid state. In contrast, in the strongly correlated regime $|G|\gg 1$, they form  
a Tonks-Girardeau gas \cite{Girardeau1960} of impenetrable hard-core particles. A characteristic feature of this regime is that the density-density correlations
\begin{equation}
g^{(2)}(z,z^{\prime})= \frac{\langle\psi^{\dagger}(z)\psi^{\dagger}(z^{\prime})\psi(z)\psi(z^{\prime})\rangle}{\langle \hat{n}(z)\rangle \langle \hat{n}(z^{\prime})\rangle}
\end{equation}
with $\hat{n}(z)=\psi^{\dagger}(z)\psi(z)$ vanish for $z=z^{\prime}$ and exhibit Friedel oscillations \cite{Friedel:2007wt} for $|z - z'|$ finite. The Friedel oscillations indicate a crystallisation of the particles by showing that they prefer to occur at specific separations from one another. For further information about the physics of interacting bosons in one dimension we here refer the reader to the review by Cazalilla et al. \cite{Cazalilla11} and now proceed to discussing realisations of this physics with interacting photons.

\subsubsection{Lieb-Liniger Model with Dark-State Polaritons}

\begin{figure}
	\centering
	\includegraphics[width=\linewidth]{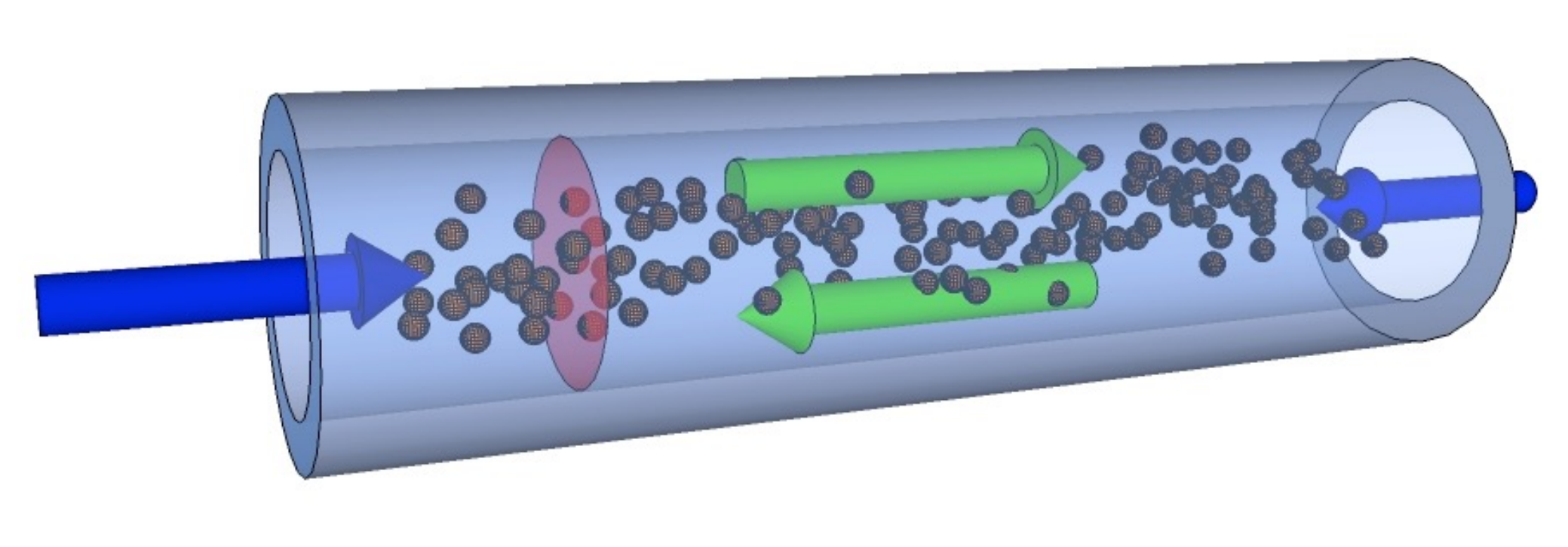}
	\caption{One dimensional waveguide, here a hollow core fibre doped with $N$ atoms that have the internal level structure depicted in figure \ref{fig:level}. Blue arrows represent classical control fields whereas green arrows represent quantum probe fields.}
	\label{fig:level-waveguide}
\end{figure}

A realisation of the model (\ref{eq:LLmodel}) with dark-state or slow-light polaritons has been proposed by Chang et al. \cite{Chang:2008qq}. The approach generalises the concept as explained in section \ref{sec:optical-BoseHubbard} to continuous models and combines it with a so called 'stationary light' regime \cite{Bajcsy:2003yq} generated by two counter-propagating control fields. A possible implementation could be a hollow-core photonic crystal fibre filled with a gas of Doppler cooled atoms \cite{Bajcsy2009}, see Fig \ref{fig:level-waveguide}, or atoms trapped in the evanescent field of an optical fibre that is tapered down to a diameter comparable or below the wavelength of the employed light \cite{Vetsch:2010uf}. 

An approach to exploiting unitary as well as dissipative interactions has been introduced by Kiffner et al.  \cite{Kiffner:2010kq} by decomposing the field $\psi$ into momentum modes.
The operator that excites a dark-state polariton in momentum $k$ is here defined as
\begin{equation}
\psi_k = \cos\theta \frac{a_{k_c + k}   + a_{-k_c + k}}{\sqrt{2}} - \frac{\sin\theta}{\sqrt{N}}\sum_{\mu=1}^{N} \sigma_\mu^{12} e^{-i k z_{\mu}}, 
\end{equation}
where $k_c$ is the momentum of the control fields, $\sin \theta =  \sqrt{N} g_{13}/\sqrt{N g_{13}^2 + 2 \Omega^2 }$, $\cos \theta =  \sqrt{2} \Omega_c/\sqrt{N g_{13}^2 + 2 \Omega^2 }$ and $N^{-1/2} \sum_{\mu=1}^{N} \sigma_\mu^{12} e^{-i k z_{\mu}}$ describes a spin coherence. The dynamics of the dark state polaritons can be described in terms of their density matrix $\rho_D$. Neglecting single particle dissipation which can be sufficiently suppressed for suitable parameters, $\rho_D$ obeys the equation of motion \cite{Kiffner:2010kq,Kiffner:2010gm},
\begin{equation} \label{eq:dissLL}
\hbar \dot{\rho}_D = -i H_{\textrm{eff}}\rho_D +i \rho_D H_{\textrm{eff}}^{\dagger} + \mathcal{I}\rho_D, 
\label{meq}
\end{equation}
where $H_{\textrm{eff}} = H_{LL}$ with effective mass $m_{\textrm{eff}}=-\hbar(N g_{13}^2 + 2 \Omega^2)/(2\delta c^2 \cos^2\theta)$ and a complex interaction constant
$\tilde{g}=2\hbar L g_{24}^2\cos^2\theta/(\Delta-\cos^2\theta \Delta\omega + i \gamma_{42}/2)$.
Here $c$ is the speed of light.
The term
\begin{equation}
\mathcal{I}\rho_D = - \textrm{Im}(\tilde{g}) \int_0^L dz \psi^2 \rho_D\psi^{\dagger 2}.
\end{equation}
describes correlated decay processes, where always two polaritons are simultaneously lost.
Equation (\ref{eq:dissLL}) thus has the form of a Lindblad master equation where the jump operator describes correlated decays of polariton pairs.

For the realisation as described by equation (\ref{eq:dissLL}), the absolute value of the Lieb-Liniger Parameter is 
\begin{equation}\label{eq:LLparamDS}
|G| = \frac{g_{13}^2 g_{24}^2 L^2 N}{c^2|\delta|\sqrt{\Delta^2 + \gamma_{42}^2/4} \, N_p}
\end{equation}
and is maximal for purely dissipative ($\Delta = 0$) interactions between the polaritons \cite{Kiffner:2010kq}, see also \cite{Hafezi:2012wq}.

The strongly correlated regime with $|G|$ larger than unity thus becomes accessible for an optical depth per atom that exceeds $160$. 
For the density-density correlations, one finds $g^{(2)}(z,z)= (1-1/N_{\textrm{ph}}^2)4\pi^2/(3 |G|^2)$ for $z=z'$ which 
vanishes in the limit  $|G|\rightarrow \infty$.
Moreover, in this strongly correlated regime, the ground state of this generalised Lieb-Liniger model is the same as in the original model with repulsive interaction \cite{Durr:2009si}, indicating a crystallisation of the polaritons.

A generalisation of the above approaches to a situation with two atomic species filling the hollow core fibre, c.f. figure \ref{fig:level-waveguide}, was considered by Angelakis et al. \cite{Angelakis2011}. These conditions give rise to two polariton species $\psi_1$ and $\psi_2$ that are each described by a Lieb-Liniger Hamiltonian as in equation (\ref{eq:LLmodel}) and in  addition are subject to a density-density interaction of the form
\begin{equation}
\int_0^L dz V_{12} \psi_1^\dag \psi_1 \psi_2^\dag \psi_2,
\end{equation}
with strength $V_{12}$ \cite{Angelakis2011}. The scenario can thus emulate Luttinger liquid behaviour and allows for exploring an analogue of spin-charge separation due to the mapping between hard-core bosons and free fermions in one dimension \cite{Girardeau1960}. Subsequently applications of this approach to simulate Cooper pairing \cite{Huo:2012rq} and relativistic field theories \cite{Angelakis:2013yo} have been investigated, see also \cite{Noh16}.

\subsubsection{Polariton correlations and dynamics}
For the above approaches to Lieb-Liniger physics with dark state polaritons, the effect of dissipative interactions and the build-up of Friedel oscillations was studied numerically in \cite{Kiffner:2011rt}.
Moreover photonic transport in a fibre as sketched in figure \ref{fig:level-waveguide} was studied in \cite{Hafezi:2011xy,Hafezi:2012wq} via a Bethe ansatz solution for the  driven Lieb-Liniger model where the setup was shown to act as a single photon switch for repulsive interactions and able to support two-photon bound states for attractive interactions. As we will discuss next, there has already been significant progress in experimental observations of the physics of such continuous one-dimensional models.

\subsubsection{Experimental Progress}
There has been substantial progress towards realising quantum many-body systems of strongly interacting photons with optical fibres, although the simulation of a Lieb-Liniger model appears to still pose challenges.
Laser-cooled $^{87}$Rb atoms have been loaded int a hollow core photonic crystal fibre by Bajcsy et al. \cite{Bajcsy2009} to demonstrate all optical switching. Via a switching beam coupled to the transition $2 \leftrightarrow 4$, see figure \ref{fig:level}, the initial transmission of a beam coupled to the transition $1 \leftrightarrow 3$ was reduced by 50\%. Another approach trapped laser-cooled neutral atoms with a multicolour evanescent field surrounding an optical nano-fibre and showed appreciable coupling of the atoms to fibre guided light modes \cite{Vetsch:2010uf}. The concept was then developed further to demonstrate switching of optical signals between two optical fibres \cite{OShea:2013si} and nanophotonic optical isolators  \cite{Sayrin:2015wt}. As achieving sufficiently strong optical nonlinearities for simulating strongly correlated quantum many-body systems with optical photons remains a challenge in these devices, atomic Rydberg media are now being considered more intensely.   

\subsection{Polaritons in ensembles of Rydberg atoms}
\label{sec:Rydberg}
For generating stronger interactions between polaritons, ensembles of Rydberg atoms have more recently received increasing attention. Here the large dipole moment of Rydberg atoms leads to large van der Waals interactions between two atoms that scale proportional to the sixth power of the principal quantum number. These interactions lead to an effect called Rydberg blockade which describes a situation where a driving field cannot generate a second Rydberg excitation in the vicinity of a Rydberg excitation due to these strong forces. Reviewing the development in this branch of research is beyond the scope of this article and we thus refer the interested reader to a series of excellent recent reviews \cite{Saffman:2010fq,Low:2012fk,Hofmann:2013uq,Chang:2014zt,Roy16} and references therein.

\section{Microwave photons in superconducting circuits}
\label{sec:microw-superc}

In the quest for realising strong effective photon-photon interactions, superconducting circuits supporting elementary excitations in the form of microwave photons have received increasing attention as they offer very favourable properties for this task. 

Importantly the wavelength of microwave photons are about $10\,$mm. Therefore resonators that trap them are of a similar size and the accuracy of available micro-fabrication techniques is sufficient for producing multiple resonators of the same resonance frequency on the same chip. More precisely, any residual disorder in resonance frequencies is well below typical values for their mutual coupling \cite{Underwood2012}. In this sense, building large, coherently coupled resonator arrays of sufficient finesse is for microwave photons significantly less challenging than in the optical domain.

Free, that is non-interacting photons are in this technology supported by various forms of LC circuits composed of an inductance and a capacitance. In these, current and voltage oscillations occur together with associated oscillations of electric and magnetic fields. For the appropriate dimensions of these circuits, their elementary excitations are microwave photons with frequencies that remain below the superconducting gap but  are large enough to keep thermal excitation numbers vanishingly small at cryogenic temperatures. 

\begin{figure}[h]
\centering
\includegraphics[width=0.8\columnwidth]{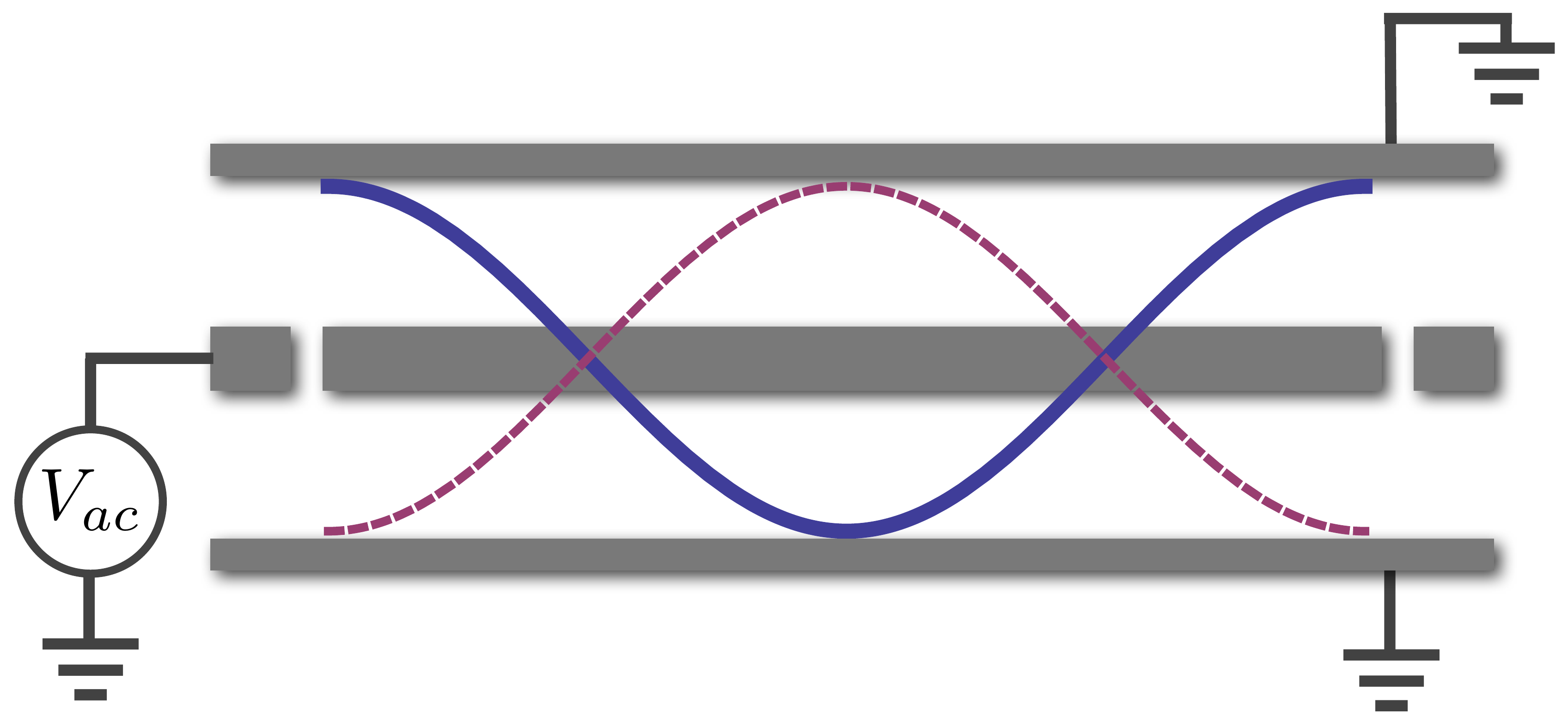}
\caption{A coplanar waveguide resonator formed by three superconducting lines on a chip. The resonator can be excited via a microwave tone applied to the central conductor, which is intersected by two capacitances that play a similar role as the mirrors in a Fabry-P\'erot cavity. The voltage profile of a resonance mode is indicated by the sinusoidal lines.}
\label{fig:striplinecavity}
\end{figure}

The employed LC circuits are thereby built as so called lumped element versions, with dimensions smaller than the wavelength of the explored photons, or as so called coplanar waveguide resonators. The latter can be viewed as a flattened version of a coaxial cable with three conductors patterned in parallel on a chip. Whereas the two outer conductors are grounded, the central conductor carries an electric signal that thus leads to a time and space dependent voltage drop --- an electromagnetic field --- between the central and outer conductors, see figure \ref{fig:striplinecavity}. The structure thus acts as a waveguide for microwave photons. A discrete spectrum of resonances can then be engineered by cutting the central conductor at two points so that capacitances form and enforce nodes of the current profile leading to anti-nodes of the voltage modes.

As LC circuits are harmonic oscillators, their excitations do not interact with each other. For applications as quantum simulators, nonlinear elements in the circuit are thus highly desirable. These are provided by Josephson junctions and the circuit elements featuring one or multiple Josephson junctions are usually called a superconducting qubits \cite{Makhlin2001,Devoret:2013qr}. There are two physical processes that determine the physics of a Josephson junction and hence a superconducting qubit, the Coulomb interaction between Cooper pairs at both sides of the junction that is determined by the junction's capacitance (including a possible shunt capacitance) and the tunnelling of Cooper pairs through the junction as quantified by the Josephson energy.

In the so called charging regime, as generated by a low capacitance junction, the energy eigenstates of the qubit are characterised by the difference in the number of Cooper pairs at both sides of the junction. Here the qubit with its strongly anharmonic spectrum can be interpreted as a matter component, playing a similar role as a two-level atom coupled to optical photons in the approaches discussed in section \ref{sec:optical-atoms}. Such scenarios have been investigated intensively following an influential experiment in 2004 by Wallraff et al.~\cite{Wallraff2004} that showed coupling between a superconducting qubit in the charging regime and a coplanar waveguide resonator, with a very high ratio of coupling strength to dissipation rates. 
Due to the analogies to cavity quantum electrodynamics (QED), this line of research was coined circuit-QED \cite{Schoelkopf2008} and initially explored scenarios with direct analogies in optical cavity QED \cite{Houck2007,Fink2008,Baur2009,Bozyigit:2011fx,Lang2011}. The strong coupling coupling strength between qubit and resonator, that is achieved in these setups, results from the large dimensions of the superconducting qubits ($\sim 1\,\mu$m) leading to large dipole moments and from the strong confinement of the electromagnetic field between the superconducting wires of the coplanar waveguide resonators. 

Superconducting qubits have been considered in various forms. Besides the regime where their eigenstates are mostly determined by the charge degree of freedom, so called phase qubits and flux qubits have been investigated intensively. Rather than discussing all these types in detail we here refer the reader to excellent reviews of this matter \cite{Makhlin2001,Devoret2004a,Clarke2008,Paraoanu:2014fk}. 

To improve robustness against dephasing noise induced by fluctuating background charges on the chip, novel designs for superconducting qubits have been developed within the last decade. The currently most prominent version is the transmon qubit \cite{Koch2007}, where the charging energy due to Coulomb interaction of Cooper pairs is reduced by shunting the junction with a large capacitance. 
Further improvement of the coherence times for transmon qubits coupled to superconducting resonators has recently been achieved by building three-dimensional resonators \cite{Paik2011a,Kirchmair:2013fk}. This design is based on a qubit only made out of two superconducting islands connected by a Josephson junction that is kept inside a three-dimensional cavity machined out of aluminium which becomes superconducting at the employed cryogenic temperatures. Experiments have already successfully explored quantum many-body physics with multiple qubits in one three-dimensional resonator, see Sec. \ref{sec:exp-3d-transmon}. 
Before reviewing quantum simulator designs in this technology we briefly discuss the quantum description of superconducting circuits.

\subsection{Quantum Theory of Circuits}
\label{sec:microw-theory}
An excellent introduction to the quantum theory of superconducting circuits has been written by Devoret \cite{Devoret1995}, so that we here merely summarise the main aspects for completeness. 

The dynamics of a superconducting electronic circuit can be described in terms of so called {\it node variables} \cite{Devoret1995}. To this end, the description is reduced to a set of independent variables by eliminating the remaining variables via Kirchhoff's laws, which state that the sum of all voltages around a loop and the sum of all currents into a node should be zeros as long as the flux through the loop and the charge at the node remain constant. A convenient choice is then to associate to each node of the network a variable $\phi(t) = \int_{-\infty}^{t} dt' V(t')$, where $V(t)$ is the voltage drop between the considered node and the ground plane. $\phi$ has dimensions of a magnetic flux and for superconducting circuits can be linked to the phase of the Cooper pair ``condensate'' $\varphi$ via the relation $\varphi= \phi / \varphi_{0}$, where $\varphi_{0} =\hbar/(2 e)$ is the reduced quantum of flux with $e$ the elementary charge. 

Using this language, the energy of an element between nodes $j$ and $j+1$ of a circuit network is given by the expressions, $\frac{\varphi_{0}^{2} C}{2} (\dot{\varphi}_{j} - \dot{\varphi}_{j+1})^{2}$ for a capacitance $C$, $\frac{\varphi_{0}^{2}}{2L} (\varphi_{j} - \varphi_{j+1})^{2}$ for an inductance $L$ and $- E_{J} \cos (\varphi_{j} - \varphi_{j+1})$ for a Josephson junction with Josephson energy $E_{J}$ (A dot denotes a time derivative.). If two identical Josephson junctions are included in a closed ring, such as in a superconducting quantum interference device (SQUID), they behave like a single effective junction for currents across the ring, where the Josephson energy of the effective junction can be tuned via a magnetic flux threaded through the ring \cite{Makhlin2001}.

A quantum theory for a circuit can be derived in the canonical way. One first sets up a Lagrangian $\mathcal{L}$ such that its Euler-Lagrange equations are identical to the classical equations of motion for the currents in the circuit. This Lagrangian is then Legendre transformed into a Hamiltonian by introducing canonical momenta $\pi_{j}$ for each node variable via $\pi_{j} = \frac{\partial \mathcal{L}}{\partial \dot{\phi}_{j}}$ \cite{Devoret1995,Nunnenkamp2011} and the theory is quantised by imposing canonical commutation relations $[\varphi_{j},\pi_{l}] = i \hbar \delta_{j,l}$.

Rather than in a chronological order, we here review the work on employing networks of superconducting circuits for the simulation of quantum many-body physics by starting with those models that have been of central interest in the optical domain as well, in particular the Bose-Hubbard and Jaynes-Cummings-Hubbard models.

\subsection{Bose-Hubbard models}
\label{sec:microw-BH}
\begin{figure*}[h]
\includegraphics[width=\textwidth]{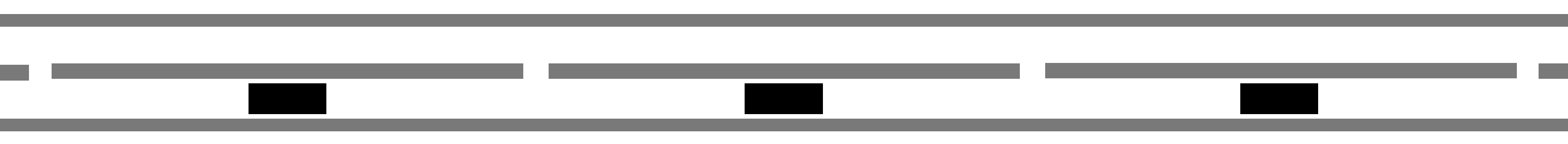}
\caption{Chain of coplanar waveguide resonators (central and grounded conductors indicated by grey lines), that each couple to a transmon qubit (balck boxes). Neighbouring resonators are coupled via a capacitance (indicated by the small gaps in the central conductor).}
\label{fig:circuitBoseHubbard}
\end{figure*}

For a lattice of coplanar waveguide resonators, that each couple to a transmon qubit, see figure \ref{fig:circuitBoseHubbard}, it can be shown that the dynamics of polaritons, formed by a superposition of resonator and qubit excitations, is described by a Bose-Hubbard Hamiltonian for suitable parameters \cite{Leib2010,Leib:2013ho}.
Since transmon qubits feature a moderate non-linearity due to their large ratio of $E_{J}/E_{C}$, where $E_C=e^2/(2C_{\Sigma})$ is the charging energy and $C_{\Sigma}$ the total capacitance of the qubit to ground, they can be modelled by a nonlinear oscillator. A single lattice site of the envisioned architecture is thus described by the Hamiltonian
\begin{eqnarray}\label{eq:TransmonQED2}
H_{\mathrm{1-site}}& =& \hbar \omega_{q} q^{\dagger}q-\frac{E_C}{12}\left(q+q^{\dagger}\right)^4 +\hbar \omega_r a^{\dagger}a \\
&+& \hbar g\left(a^{\dagger}q+aq^{\dagger}\right) \nonumber
\end{eqnarray}
where $a^{\dag}(a)$ and $q^{\dag}(q)$ are creation (annihilation) operators of the resonator and qubit and $\omega_{q} = \hbar^{-1} \sqrt{8E_CE_J}$ is the transition 
frequency of the qubit. The coupling $g$ between resonator and qubit typically greatly exceeds the dissipation rates for both modes. If moreover, this coupling is stronger than the nonlinearity of the transmon qubits, $\hbar g > E_{C}$ the qubit and resonator modes hybridise and two species of polaritons with annihilation operators $p_+$ and $p_-$ of the same form as in equation (\ref{eq:PolaritonModesBH})
become the elementary excitations of the system \cite{Leib2010}, see also section \ref{sec:exciton-pol}.

Neighbouring resonators can for example be coupled via a mutual capacitance $C_{J}$, see figure \ref{fig:circuitBoseHubbard}, leading to the energy,
\begin{equation}\label{eq:resonator-coupling}
\frac{C_{J}}{2} \Delta V_{j,j+1}^{2} = \hbar J_{a} [(a_{j} + a_{j}^{\dag}) - (a_{j+1} + a_{j+1}^{\dag})]^{2}
\end{equation}
where $\Delta V_{j,j+1}$ is the voltage difference across the coupling capacitance. In a rotating wave approximation, this coupling leads to frequency shifts for both coupled resonators and a tunnelling of photons between both resonators at a rate $J_{a}$.

In terms of the polaritons $p_+$ and $p_-$, c.f. equation (\ref{eq:PolaritonModesBH}), the Hamiltonian describing the circuit reads
\begin{equation} \label{eq:circuitBHHamiltonian}
H = H_{p_+}+H_{p_-}+H_{dd},
\end{equation}
where $H_{p_{\pm}}$ are Bose-Hubbard Hamiltonians as in equation (\ref{eq:bosehubbard}) with interactions $U_{+} = - E_{C} \sin^{4}(\theta)$ and $U_{-} = - E_{C} \cos^{4}(\theta)$,
polariton tunnelling $J_{+} = J_{a} \sin^{2}(\theta)$ and $J_{-} = J_{a} \cos^{2}(\theta)$. $H_{dd} = U_{+-} \sum_{j} p_{+,j}^{\dag}p_{+,j} p_{-,j}^{\dag}p_{-,j}$ describes a density-density interaction between both species with $U_{+-} = E_{C} \sin^{2}(\theta) \cos^{2}(\theta)$. In the derivation of equation (\ref{eq:circuitBHHamiltonian}) the difference between the frequencies of both polariton species has been assumed to greatly exceed all interaction strength and tunnelling rates so that all processes that would lead to a mixing of both species are strongly suppressed, and a rotating wave approximation has been applied.

\begin{figure}[h]
\includegraphics[width=\columnwidth]{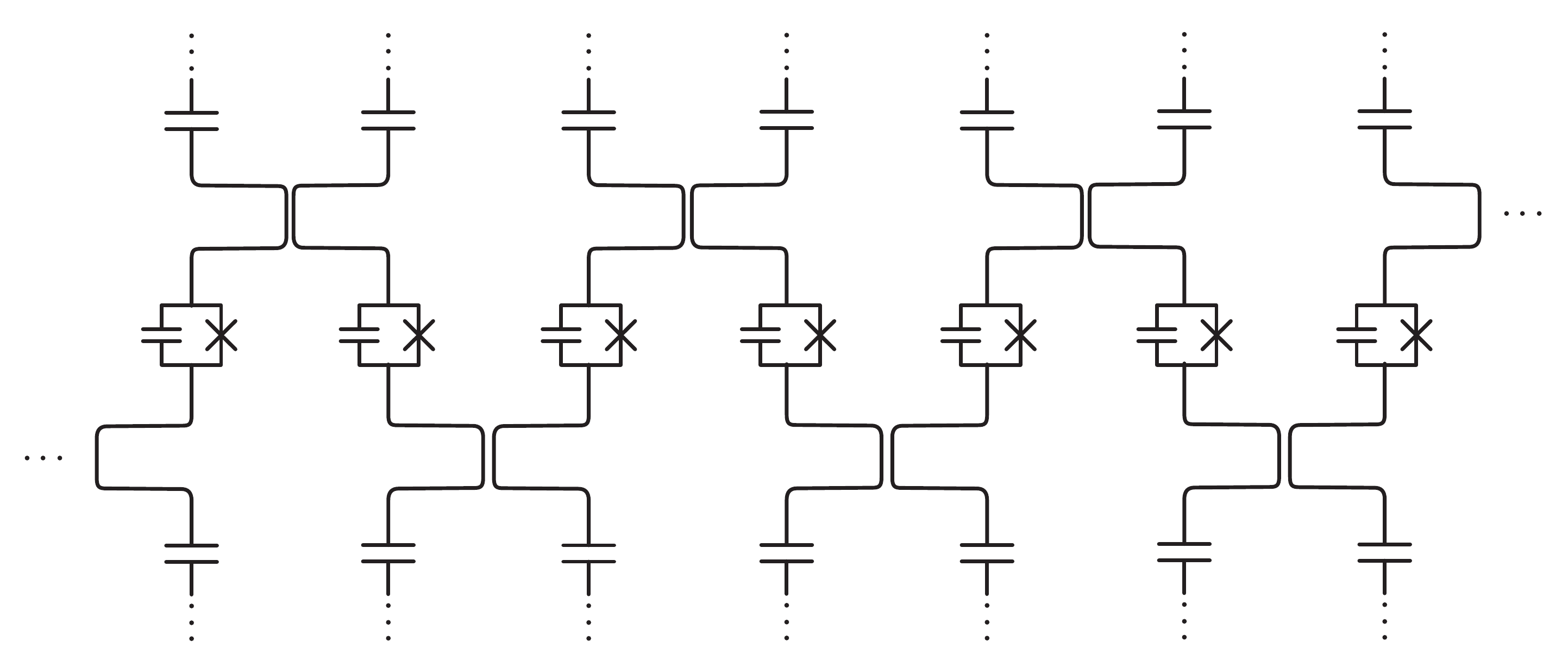}
\caption{Sketch of array of coplanar waveguide resonators (only central conductors are drawn), that are nonlinear due to a Josephson junction (shunted by a capacitance) inserted into their central conductor. All resonators couple to in- and output lines on the top or bottom. The mutual coupling between the resonators is generated by bringing them into close proximity at suitable points so that a mutual capacitance and inductance forms \cite{Baust:2015cl}. Figure reproduced from  \cite{Leib:2012db}.}
\label{fig:nonlinresarray}
\end{figure}

Another approach to Bose-Hubbard physics in superconducting circuits \cite{Leib:2012db} considers coplanar waveguide resonators that have been made nonlinear due to Josephson junctions or dc-SQUIDs inserted in their central conductors at the location of a current node of the bare resonator \cite{Bourassa:2009uq,Leib:2012db}, see figure \ref{fig:nonlinresarray} for a sketch. This approach leads to a Bose-Hubbard model for a single excitation species. The approach can also be interpreted as the SQUIDs and resonators being ultra-strongly coupled \cite{Deppe2008,Bourassa:2009uq,Leib:2012db,Bourassa:2012fk} so that the splitting between the two normal modes in each resonator becomes comparable to their frequencies and mixing processes between excitations in both normal modes are truly absent. 

The concept of making a coplanar waveguide resonator nonlinear by inserting a Josephson junction or SQUID into its central conductor at the location of a voltage node or current anti-node can be taken a step further by inserting multiple junctions into the central conductor. When applied to a waveguide or very long resonator, this approach gives rise to an extended ``artificial'' medium, which is optically nonlinear at the single photon level. The idea has been considered by Leib et al. \cite{Leib:2014qd}, where a synchronised switching of the architecture's normal modes from a low excitation quantum regime to a highly excited classical regime has been found.

\subsubsection{Non-local interactions}
\label{sec:nonlocalints}
\begin{figure}
\centering
\includegraphics[width=0.8\columnwidth]{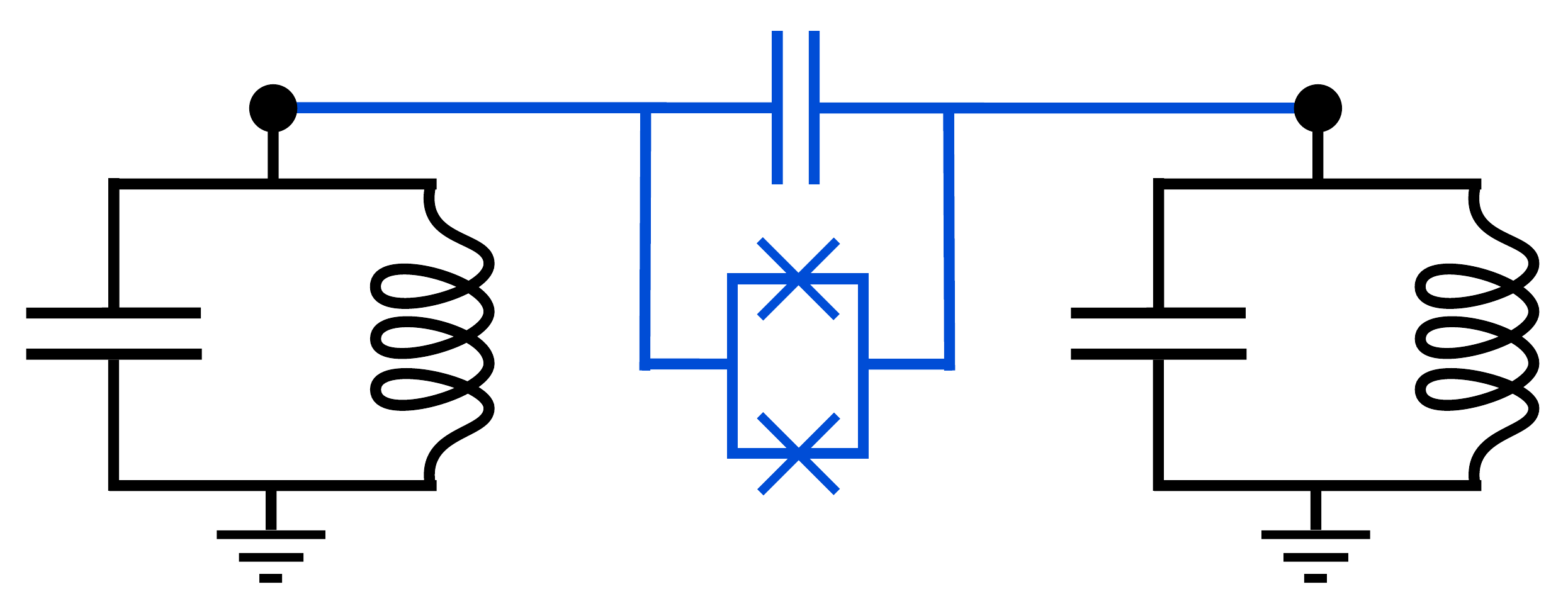}
\caption{Circuit for a nonlinear coupler that gives rise to cross Kerr interactions as specified in equation (\ref{eq:crossKerr}).}
\label{fig:crossKerr}
\end{figure}

In contrast to optical photons interacting with solid state emitters or atoms, microwave photons in superconducting circuits can be made to interact non-locally \cite{Jin:2013qp}. That is a photon in one resonator can scatter off a photon in a second, even spatially distant, resonator. Such processes are described by cross-Kerr interactions of the form
\begin{equation} \label{eq:crossKerr}
V \, a_1^\dag a_1 \, a_2^\dag a_2,
\end{equation}
where $a_1 (a_2)$ annihilates a photon in resonator 1(2). The circuit that generates this type of interactions together with correlated tunnelling processes is sketched in figure \ref{fig:crossKerr}.
The dc-SQUID connecting the two resonators gives rise to an energy contribution of $- E_J \cos(\varphi_1 - \varphi_2)$, where $\varphi_{1(2)} \propto a_{1(2)} + a_{1(2)}^\dag$, and an expansion up to 4-th order in $\varphi_1 - \varphi_2$ leads to
\begin{eqnarray} \label{eq:crossKerr2}
H_{cK} & = & \hbar \tilde{\alpha} \omega (a_1^\dag a_2 + a_1 a_2^\dag) - 2 \alpha E_C a_1^{\dag}a_1 a_{2}^{\dag}a_{2} \\
& + & \alpha E_C\big(a_ia_{j}^{\dag}a_{j}^{\dag}a_{j} + a_i^{\dag}a_i^{\dag}a_ia_{j} 
-\frac{1}{2}a_i^{\dag}a_i^{\dag}a_{j}a_{j}+ \textrm{H.c.} \big), \nonumber
 \end{eqnarray}
where $E_C = e^2/[2(C+2C_J)]$ and $\alpha = C_J/(C+2C_J)$ with $C$ the capacitance in each resonator and $C_J$ the capacitance that shunts the dc-SQUID. The coefficient $\tilde{\alpha}$ can be made vanishingly small by tuning the SQUID to the point, where photon tunnelling via the SQUID and via the shunt capacitance interfere destructively and cancel each other \cite{Neumeier:2013bq}. When all couplings between neighbouring resonators in a large lattice are built as in figure \ref{fig:crossKerr}, one arrives at a Bose-Hubbard Hamiltonian augmented by the cross-Kerr interactions (\ref{eq:crossKerr}) and correlated tunnelling processes \cite{Jin:2013qp}. This scenario can give rise to a density wave type ordering, see also section \ref{sec:phase-diag}.

Lattice elements coupled by Josephson junctions have furthermore been considered for the simulation of Anderson and Kondo lattices \cite{Garcia-Ripoll:2008ta} and tunable coupling elements \cite{Baust:2015cl}. 

\subsubsection{Relation to Josephson junction arrays}
Bose-Hubbard physics in superconducting architectures has been investigated in the early 1990 already, well before the realisations of the model in optical lattices. The investigated structures consisted of Josephson junction arrays where Cooper pairs could tunnel between superconducting islands through the junctions and interact via Coulomb forces on each island, see \cite{Fazio2001} for a review. The practical difference of the new generation of networks discussed here is that the individual network nodes are separated by larger distances on the chip and can therefore be individually addressed via control lines. The high precision control over the individual nodes allows to suppress disorder much better than in Josephson junction arrays, where it was a significant limitation to experiments. Yet circuit-QED lattices also allow to emulate further many-body models, such as the Jaynes-Cummings-Hubbard model.

\subsection{Jaynes-Cummings-Hubbard models}
\label{sec:microw-JCH}
If the nonlinearities of the employed qubits are larger than all other interaction and tunnelling processes in the circuit network, its dynamics can no longer be approximated by a Bose-Hubbard model for polaritonic excitations, but is described by a Jaynes-Cummings-Hubbard model as given in equation (\ref{eq:JCHubbard}) \cite{Schmidt2010a,Houck2012}.
In this regime, $hg \ll E_{C}$ and the qubit is approximated by a two-level system, $b \to \sigma^-$ and $b^\dag \to \sigma^+$ in equation (\ref{eq:TransmonQED2}), which gives rise to a Jaynes-Cummings model describing the individual lattice site. Together with the tunnelling of microwave photons between adjacent resonators as described in equation (\ref{eq:resonator-coupling}), this leads to a Jaynes-Cummings Hubbard model for the entire lattice. 

The approximation of a circuit QED system consisting of a transmon qubit and a resonator by a Jaynes-Cummings Hamiltonian has been investigated experimentally, where it was possible to show the characteristic scaling of its nonlinearity with $\sqrt{n}$, where $n$ denotes the number of excitations in the system \cite{Fink2008}.
In contrast to atoms, the approximation of a superconducting qubit as a two level system needs to be considered with much more care, in particular for transmon qubits, where the increased robustness against charge noise comes at the expense of a slightly reduced nonlinearity as compared to charge qubits. Indeed the experiment \cite{Lang2011} also found corrections to the approximations by a two-level system due to the finite nonlinearity of the qubit.

A Kagom\'e lattice of coupled coplanar waveguide resonators that each interact with a superconducting qubit such that individual lattice sites can be described by a Jaynes-Cummings model has been investigated by Koch et al. \cite{Koch2010,Houck2012}. The approach approach considered three-port coupling elements that are capacitively connected to the three coplanar waveguide resonators that meet at each vertex. These couplers lead to time-reversal symmetry breaking, which is a prerequisite for accessing certain classes of quantum many-body states such as fractional quantum Hall states. We discuss quantum simulators for this class of systems in more detail in section \ref{sec:artificial-gauge}.

A dimer of two capacitively coupled coplanar waveguide resonators that are each capacitively coupled to a superconducting qubit has been investigated by Schmidt et al. \cite{Schmidt2010a} to explore a localisation-delocalisation transition from a self-trapped phase to an oscillating phase as the interaction strength between qubits and cavity modes crosses a critical value. Discrepancies between the classical and quantum prediction for this critical value have been resolved in \cite{Shapourian:2016fk}. This transition has later been observed in an experiment by Raftery et al. \cite{Raftery:2014fj}, c.f. Sec. \ref{sec:self-trap}.  

Moreover approaches to simulating spin systems in coupled arrays of circuit cavities and superconducting qubits have been put forward recently \cite{Kurcz:2014rp,Ashhab:2014it}.

\subsection{Digital Quantum Simulations}
\label{sec:microw-DQS}
In this review we focus our attention on quantum simulations where an experimentally well controllable system is tuned such that it emulates a specific Hamiltonian. This approach to quantum simulation is termed {\it analog} quantum simulation. In contrast, one can also employ a device where specific unitary operations can be performed with high precision. A sequence of such operations can then lead to the same dynamics as the target Hamiltonian of a quantum simulation as can be seen via Trotter's formula. This approach of {\it digital} quantum simulation is well amenable to devices that have been designed to implement quantum gates and was recently demonstrated in superconducting circuits to simulate spin \cite{Salathe:2015yf} and fermionic \cite{Barends:2015dz} Hamiltonians in one dimension as well as molecular energies \cite{OMalley15}. 
The theory for these approaches was worked out in \cite{Heras:2014zv} and \cite{Aspuru-Guzik:2005yu}, see also \cite{Mezzacapo:2014wt} for a related theoretical and \cite{Barends16} for related experimental work.
Other approaches considered the quantum simulation of the regime of s so called ultra-strong light-matter coupling \cite{Deppe2008,Ridolfo:2012lj,Ridolfo:2013vk,Stassi:2013yp} in driven systems \cite{Ballester:2012zh}.

\subsection{Experimental Progress}
\label{sec:microw-exp}
The experimental progress towards assembling and controlling larger and larger networks of superconducting circuits has been remarkable in recent years. Whereas most of the efforts are aiming at implementations of quantum information processing tasks, quantum simulation applications have received increasing interested lately. 

For quantum computation applications, three-qubit gates have been demonstrated in 2012 \cite{Reed2012,Lucero2012,Fedorov:2012ef}. More recently, larger networks of coherently coupled qubits (up to 9 at the time of writing) with performance fidelities suitable for surface code computation \cite{Barends:2014fk} and state preservation by error detection in the repetition code \cite{Kelly:2015vn} have been shown. To boost the scalability of such networks, multilayer structures are now being built \cite{Brecht:2016gf}. 

For quantum simulation of quantum many-body systems, very low disorder (below $10^{-4}$) has been shown for 12 coupled coplanar waveguide resonators on a Kagom\'e lattice \cite{Underwood2012} and large resonator lattices (more than 100 resonators) have been built \cite{Houck2012}. Weak localisation has been simulated in a multiple-element superconducting quantum circuit \cite{Chen:2014uq} and topological phases together with transitions between them have been measured via the deflection of quantum trajectories with two interacting qubits \cite{Roushan:2014kx}. Moreover digital quantum simulations of spin models including their mapping to non-interacting fermions \cite{Salathe:2015yf,Barends:2015dz} have been shown and a novel quantum simulation concept employing an experimental generation of Matrix Product States \cite{Orus:2014ul} has been demonstrated for the Lieb-Liniger model \cite{Eichler:2015sf}. There also have been first few-site realisations of analogue quantum simulation devices for Bose-Hubbard chains which may form building blocks of larger scale quantum simulators with interacting microwave photons. 

\subsubsection{Few-site Bose-Hubbard chains}
\label{sec:exp-3d-transmon}
A dimer of two lumped element resonators has been empoyed by Eichler et al. to show quantum-limited amplification and entanglement \cite{Eichler:2014vn}.
In their setup, both resonators are nonlinear because their inductors are formed by a series of dc-SQUIDs and are mutually coupled via a capacitance.
Here a moderate nonlinearity is desirable to allow for an appreciable bandwidth of the amplification mechanism.

A chain of three capacitively coupled transmon qubits has been realised inside a microwave cavity by Hacohen-Gourgy et al. \cite{Hacohen-Gourgy:2015kx}.
Here all qubits couple dispersively to a common resonance mode of the cavity and engineered cooling and heating processes are generated by driving the cavity resonance with red or blue detuned input tones. These processes act as a quantum bath that can exchange energy and entropy with the Bose-Hubbard chain but conserves its excitation number,  see also Sec. \ref{sec:microw-chem-pot}. An extension to more qubits in a common resonator for exploring dipolar spin models has been considered in \cite{Dalmonte:2015uq}.

\section{Quantum many-body dynamics and phase diagrams}
\label{sec:phase-diag}

Besides their application as quantum simulators, the approaches to generate effective quantum many-body Hamiltonians discussed in this review, also call for investigations of the many-body phenomena they give rise to. An important question is whether the phase diagrams for the approaches as discussed in sections \ref{sec:optical-atoms} and \ref{sec:microw-superc} show any deviations from those of the target models or what the phase diagrams of the novel models motivated by these approaches are? We discuss these questions in this section, see also related reviews by Tomadin et al. \cite{Tomadin2010}, Schmidt et al. \cite{Schmidt:2013bv} and Le Hur et al. \cite{LeHur15}. As most of the interest was initially focussed on equilibrium regimes, we begin our discussion with these. 

\subsection{Equilibrium studies}
\label{sec:phase-equi}
Initial investigations of the phase diagrams of effective Hamiltonians for interacting photons or polaritons in arrays of coupled cavities \cite{Greentree2006,Rossini2007,Koch2009a} considered equilibrium scenarios by introducing a chemical potential, the physical realisation of which remains an open question, see section \ref{sec:microw-chem-pot} for further discussion of this aspect. 
For the Jaynes-Cummings-Hubbard model, these works addressed the immediate question whether the phase diagram would show any differences to the Bose-Hubbard case due to the microscopic differences related to the two component nature of the model with atomic excitations and photons. Before discussing the phase-diagrams of the most prominent models, we here first elaborate further on the absence of a chemical potential for photons and approaches to create such a potential. 

\subsubsection{Engineering a chemical potential for photons}
\label{sec:microw-chem-pot}
Photons can typically only be trapped for limited times in optical or microwave resonators. The achievable trapping times are moreover comparable or even below the experimental equilibration time-scales. In addition, the interaction of photons with matter involves absorbtion and emission processes. As a consequence of these properties, the number of photons is usually not conserved during an experiment and there is no equivalent to a chemical potential for photons. 

In some experimental situations thermalisation of photons via scattering with phonons or repeated absorption-emission cycles was achieved on sufficiently fast time-scales so that a quasi-equilibrium builds up. Prominent examples for such situations are the condensation of exciton polaritons \cite{Kasprzak,Keeling:2007kx,Kirton:2013rm,Leeuw:2013eu,Sobyanin:2013yg} in a semiconductor microcavity or the number-conserving thermalisation and Bose-Einstein condensation of a two-dimensional photon gas in a dye-filled optical microcavity \cite{Klaers}. Despite the observation of the characteristic features of Bose-Einstein condensation these non-equilibrium cases are distinct from the equilibrium situation \cite{Yukalov:2012sf} and creating a proper chemical potential for light remains an open question.

Concepts for generating such a chemical potential for photons have been proposed based on a parametric modulation of the system's coupling to its environment at a frequency that exceeds the highest spectral component of the environment \cite{Hafezi:2015rt}. Moreover, connections to regimes of ultra-strong light matter coulping have been discussed as their ground states can be viewed as containing a non-vanishing photon number which is however not observable without further manipulation
 \cite{Schiro2012,Hafezi:2015rt}. As the generation of a chemical potential for photons remains a challenging task, most discussions of equilibrium phase diagrams simply introduced such a potential by hand. We start our discussion of these works by reviewing those based on  mean-field approaches that are expected to be accurate in high dimensional lattices.

\subsubsection{Mean-field calculations for three dimensions}
\begin{figure}[h]
\includegraphics[width=\columnwidth]{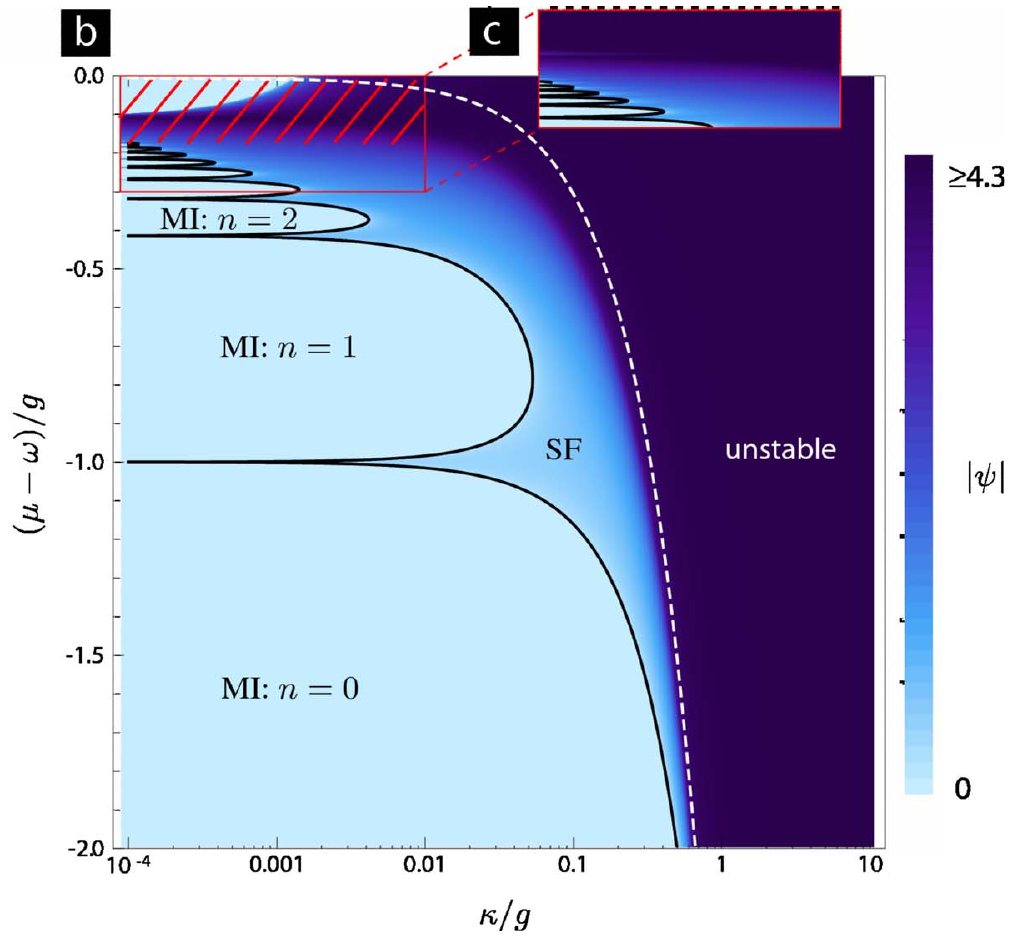}
\caption{Ground state phase diagram of the Jaynes-Cummings-Hubbard model for $\Delta = 0$ as obtained from equation (\ref{eq:JCHmeanfield}) for a three dimensional lattice with $z=6$. Note that the axes labels use a different notation with $\beta$ instead of $g$ and $\kappa$ instead of $J_{JC}$. Figure reproduced from \cite{Koch2009a}}
\label{fig:JCHgroundstate}
\end{figure}

The first study by Greentree et al. \cite{Greentree2006} used a mean-field approach that is expected to become increasingly accurate with higher (three or more) lattice dimensions. Introducing $\psi = \langle a_j \rangle$ as a superfluid order parameter, a mean-field decoupling was performed in the Hamiltonian (\ref{eq:JCHubbard}) and a chemical potential $\mu$ was added to get
\begin{eqnarray} \label{eq:JCHmeanfield}
\tilde{H} & = & \sum_{\vec{R}} H^{JC}_{\vec{R}} -\mu \sum_{\vec{R}} \left(a_{\vec{R}}^{\dagger} a_{\vec{R}} + \sigma_{\vec{R}}^{+} \sigma_{\vec{R}} \right) \\
& - & z J_{JC} \sum_{\vec{R}} \left( a_{\vec{R}}^{\dagger} \psi + \textrm{H.c.} \right) + z J_{JC} |\psi|^2, \nonumber
\end{eqnarray}
where $z$ denotes the number of neighbouring lattice sites, i.e. $z=6$ for three dimensions. Depending on whether the value of $\psi$ that minimises the expectation value of $\tilde{H}$ vanishes or not, the system is in a Mott insulating state or supports a super-fluid component. The resulting phase diagram for the case where the transition frequency of the two-level system $\omega_{0}$ equals the cavity resonance frequency $\omega_{C}$, $\Delta = \omega_{0}-\omega_{C} = 0$, is reproduced in figure \ref{fig:JCHgroundstate}.  The analogous phase diagram for the Bose-Hubbard model had already been calculated in the late 1980s by Fisher et al. \cite{Fisher89}. 
Using a field-theoretic approach, Koch and Le Hur showed that the Jaynes-Cummings-Hubbard and Bose-Hubbard models are in the same universality class and that Jaynes-Cummings-Hubbard features multicritical curves, which parallel the presence of multicritical points in the Bose-Hubbard model \cite{Koch2009a}.

Comparisons between the mean-field results and exact calculations for small lattices have found that the Mott lobes shrink in finite systems, similar to the Bose-Hubbard model \cite{Makin2008}.
Moreover, the phase diagram of the Jaynes-Cummings-Hubbard model changes substantially for ultra-strong light matter coupling, where it maps to the transverse field Ising model and features a discrete parity symmetry-breaking transition \cite{Schiro2012}.

\subsubsection{Calculations for one and two dimensions}

\begin{figure}[h]
\includegraphics[width=\columnwidth]{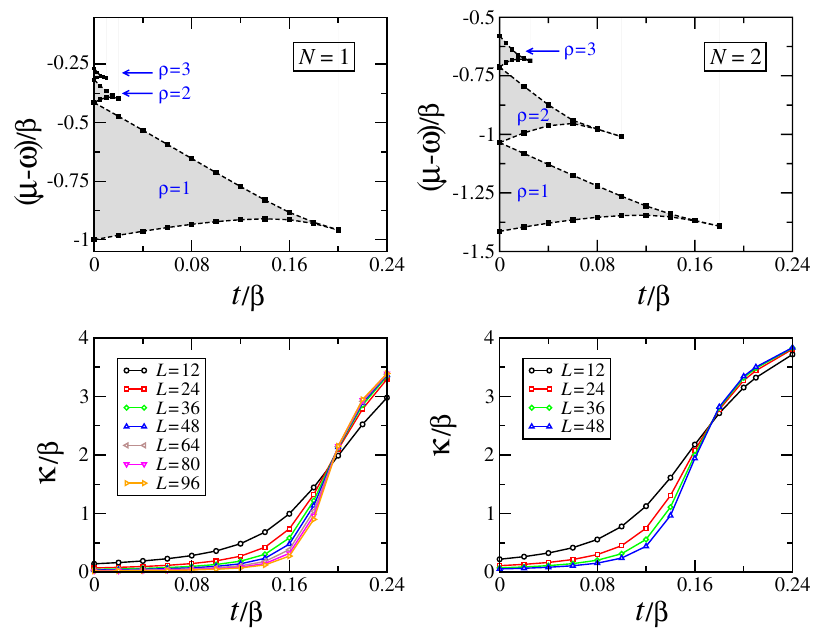}
\caption{Ground state phase diagram of the Jaynes-Cummings-Hubbard model for $\Delta = 0$ for a one-dimensional lattice. Note that the axes labels use a different notation with $\beta$ instead of $g$ and $t$ instead of $J_{JC}$. Left (right) column for $N=1$ $(N=2)$ atoms per cavity. The bottom row shows the compressibility $\kappa = \partial n / \partial \mu$, where $n$ is the average number of excitations per lattice site. Figure reproduced from \cite{Rossini2007}.}
\label{fig:JCHubbard1d}
\end{figure}

Whereas mean-field approaches are expected to become increasingly accurate for higher and higher lattice dimensions, ground states and dynamics of one-dimensional systems can often be efficiently calculated using numerical approaches based on the Density Matrix Renormalisation Group (DMRG) \cite{Schollwock:2011ys}. For the approaches described in sections \ref{sec:optical-BoseHubbard} and \ref{sec:optical-JCHubbard} the ground state phase diagrams (after a chemical potential term had been added) have been obtained with these techniques by Rossini et al. \cite{Rossini2007,Rossini2008}, see figure \ref{fig:JCHubbard1d}. Here, the microscopic origin of the nonlinearity in the explicit form of the atomic level structure has been taken into account for the approach to the Bose-Hubbard model, c.f. section \ref{sec:optical-BoseHubbard}, and deviations in the phase-diagram of the model from the original Bose-Hubbard model, c.f. equation (\ref{eq:bosehubbard}), have been found due to small atom numbers. Moreover, the existence of a glassy phase due to non-vanishing disorder in the number of atoms per cavity has been predicted.

Phase diagrams for one and two-dimensional systems have also been calculated with a variational cluster approach \cite{Aichhorn2008,Knap,Knap2011} or quantum Monte Carlo calculations \cite{Pippan,Hohenadler2011,Hohenadler2012}, and analysed differences between the Bose-Hubbard and Jaynes-Cuminngs-Hubbard models due to the composite nature of the excitations in the latter. An analytic strong-coupling theory based on a linked-cluster expansion for the phase diagram of the Jaynes-Cummings-Hubbard model and its elementary excitations in the Mott phase has been derived by Schmidt et al. \cite{Schmidt2009,Schmidt2010}.

Recent work has also predicted quantum phase transitions in finite size and even single site systems of Rabi \cite{Hwang:2015lj} and Jaynes-Cummings models \cite{Hwang16}.

\subsection{Non-equilibrium explorations}
\label{sec:non-equi}
In contrast to ultra-cold atoms, photons or polaritons are only trapped for considerably shorter times in the samples.
On the other hand photons can be produced and injected into the device at low 'cost' and in large numbers. As moreover a chemical potential for photons does not appear in nature but needs to be carefully engineered, c.f. section \ref{sec:microw-chem-pot}, it appears to be much more natural and feasible to explore quantum many-body systems of interacting photons in driven scenarios where continuous or pulsed inputs counteract dissipation. We first consider pulsed input drives and discuss continuous driving schemes afterwards.

\subsubsection{Pulsed driving fields}
Properties of the ground state phase diagrams of Bose-Hubbard and Jaynes-Cummings-Hubbard models can be investigated in non-equilibrium states using the following concept. The system parameters are initially tuned such that photon tunnelling between resonators is strongly suppressed and on-site interactions are very strong. Due to the nonlinearity of their spectra, an input pulse can generate a single excitation in each lattice site and thus prepare the system in a state very similar to a Mott insulating state. Despite not being the ground state, this initial state is an approximate eigenstate of the system for the chosen parameters. By changing the parameters of the system slowly enough to generate an adiabatic sweep, the system will stay in this energy eigenstate, which will however change its character and become very similar to the superfluid ground state when the tunnelling is increased and nonlinearities are decreased. This sequence has been analysed by Hartmann et al. \cite{Hartmann06} in terms of the fluctuations of the polariton number in each lattice site and by Angelakis et al. \cite{Angelakis2007} for the excitation number fluctuations in a Jaynes-Cummings lattice. Particle number fluctuations are strongly suppressed in the Mott insulating regime where particles are localised but become finite in the superfluid regime where the particles become delocalised.

A scenario similar to sudden quenches has been explored by Tomadin et al. \cite{Tomadin2010a}, where, independently of the system parameters, the cavity array was assumed to be initialised into a direct product of single excitation Fock states for each cavity by a short but intense input pulse. By exploring the subsequent non-equilibrium dynamics, it was found that the rescaled superfluid order parameter $\psi/\sqrt{n}$ ($n$ is the excitation density) decays to zero in the Mott insulating but remains finite in the superfluid regimes, see also \cite{Creatore:2014pd} for analogue results for lattices with disorder. Due to the possibilities of local and single-site control offered by many setups for resonator arrays, one can also explore quenches that are not uniform across the lattice. For example if a bipartite lattice is initially prepared in a superfluid regime in one half and a Mott insulating regime in the other half, all excitations migrate to the superfluid half upon switching on a small tunneling between both parts \cite{Hartmann:2008wo}. This example shows that the tendency of non-equilibrium quantum systems to explore all the accessible Hilbert space,
which typically results in the formation of local equilibria \cite{Eisert:2015vc}, can lead to states which are strongly inhomogeneous and not translation invariant provided the underlying system breaks such translation invariance.

\subsubsection{Continuous driving fields and driven-dissipative regimes}
\label{sec:phase-drivendiss}
Due to inevitable experimental imperfections, photons dissipate after a relatively short time from the quantum simulation devices described in this review. These photon losses can be compensated for by continuously loading new photons into the device via coherent or incoherent input fields. This approach is very feasible as photons are much easier and cheaper to produce as compared to for example ultra-cold atoms \cite{Diehl,Diehl2}. The emulated quantum many-body systems are therefore most naturally and feasibly explored in driven-dissipative regimes where input drives continuously replace the dissipated excitations and the dynamical balance of loading and loss processes eventually leads to stationary states.

This mode of operation should not be viewed as merely a means of compensating for an imperfection of the technology. In fact, quantum many-body systems are much less explored in such non-equilibrium scenarios than in equilibrium regimes. Investigating driven-dissipative regimes of interacting photons thus leads onto largely unexplored territory and may lead to interesting discoveries. One may even search for non-equilibrium phase transitions, that is for points where the properties of the stationary state of some driven-dissipative dynamics change abruptly as one of the system's parameters is varied \cite{Kessler2012a}.

For investigations of driven-dissipative regimes, coherent driving fields at each lattice site are often considered. To be able to perform calculations with a time-independent Hamiltonian, it is often useful to move to a frame that rotates at the frequencies $\omega_d$ of the input fields. In this frame, the frequency of a photon in each cavity (resonator), $\omega_r$, is replaced by the detuning between $\omega_r$ and the frequency of the drive $\omega_r \to \Delta_r = \omega_r - \omega_d$ and  similarly the transition frequency of emitters in the resonators is replaced by their detuning from the drive frequency,  $\omega_e \to \Delta_e = \omega_e - \omega_d$. A field that continuously drives the cavity modes is, in this rotating frame, described by an additional term in the Hamiltonian,
\begin{equation} \label{eq:driving-terms}
H_d = \sum_j \left(\frac{\Omega_j}{2}e^{-i\varphi_j} a_j^\dag + \textrm{H.c.}\right),
\end{equation}
where $\Omega_j$ is the amplitude and $\varphi_j$ the phase of the driving field at lattice site $j$. Note that while a global phase of all diving fields can be gauged away, the assumption of coherent drives requires choosing a relative phase between each pair of drives.
For the dissipation, local particle losses are typically assumed and modelled by Lindblad type damping terms \cite{Breuer1997}. Hence, the driven-dissipative models discussed here are described by master equations of the from
\begin{equation} \label{eq:master}
\dot{\rho} = -\frac{i}{\hbar} \left[ H , \rho \right] + \sum_j \frac{\gamma}{2} \left(2a_j \rho a_j^\dag - a_j^\dag a_j \rho - \rho a_j^\dag a_j \right),
\end{equation}
where $\gamma$ is the rate at which photons are lost from the device and $H$ the Hamiltonian of the considered model including driving terms such as in equation (\ref{eq:driving-terms}) and written in a frame that rotates at the frequencies of the driving fields. Note that the form of the damping terms is invariant under the transformation into this rotating frame. In our discussion of stationary regimes of driven-dissipative quantum many-body models, we first consider the Bose-Hubbard model.

\subsubsection{Driven-dissipative Bose-Hubbard model}
The driven-dissipative regime of a single resonator that features a Kerr nonlinearity for the cavity mode and thus corresponds to a single lattice site of the Hamiltonian (\ref{eq:bosehubbard}) has been studied by Drummond and Walls \cite{Drummond:1980dp} long before quantum simulation applications have been considered. These initial studies aimed to explore the physics of nonlinear polarisability and where later extended to consider solid-state nanostructures for single photon sources \cite{Ferretti:2012sp}.

\begin{figure}
\includegraphics[width=\columnwidth]{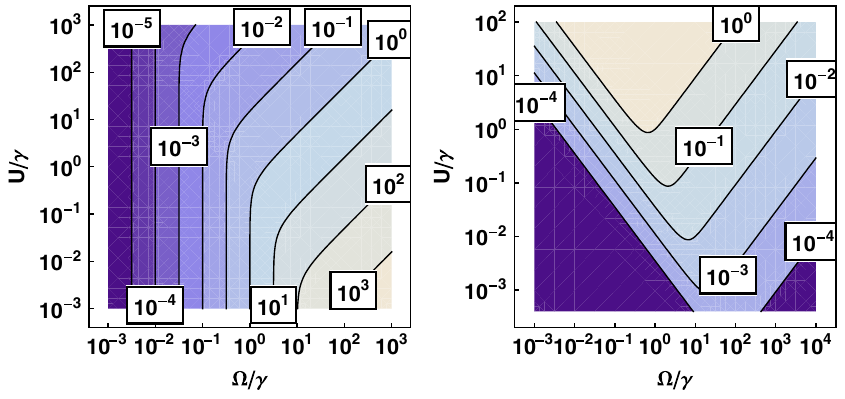}
\caption{Density of photons in the classical background (left) and ratio of photon densities in the quantum fluctuations and classical background (right), for relative phase of $\pi/2$ between adjacent driving fields. A semiclassical description is justified for small values in the right plot. Reproduced from \cite{Hartmann10}.}
\label{fig:ddBHsemiclass}
\end{figure}

For a Bose-Hubbard model describing tunnel-coupled resonators, that is driven by coherent input fields at all lattice sites, one would expect that the field in the cavity array should inherit the classical and coherence properties of the driving fields provided their intensity is strong enough to dominate over the effects caused by the nonlinearities. The boundaries of this semi-classical regime have been calculated by Hartmann \cite{Hartmann10} via a linearisation in the quantum fluctuations around the classical background component, see figure \ref{fig:ddBHsemiclass}.

In the regime of strong interactions or nonlinearities, the number of excitations per lattice site is at most $1/2$. This bound becomes obvious in the limit of very strong interactions, where each lattice site can be approximated by a two-level system as higher excited states are far off resonance to the drive. Yet, as a coherent field cannot generate inversion \cite{Wallsbook} these two-level systems have an excitation probability below $1/2$. Consequently, Mott insulating regimes with commensurate filling can in this way not be generated.

A lattice system with low particle density, more precisely a system where the average inter particle spacing greatly exceeds the lattice constant, can be viewed as an approximation to a continuum system. The properties of a one-dimensional driven-dissipative Bose-Hubbard model in the strongly interacting regime should thus rather be compared to a Lieb-Liniger model, c.f. equation (\ref{eq:LLmodel}). 

To explore whether Lieb-Liniger physics can be observed in such driven-dissipative regimes, Carusotto et al. \cite{Carusotto2009} considered a five-site version of the Hamiltonian (\ref{eq:bosehubbard}) and investigated whether the characteristic energies of the collective strongly correlated many-body states could be seen in a spectroscopy analysis. They scanned the frequency of the driving fields through the relevant range and found resonance peaks at the transition-frequencies of the Hamiltonian (\ref{eq:bosehubbard}).

\begin{figure}
\includegraphics[width=\columnwidth]{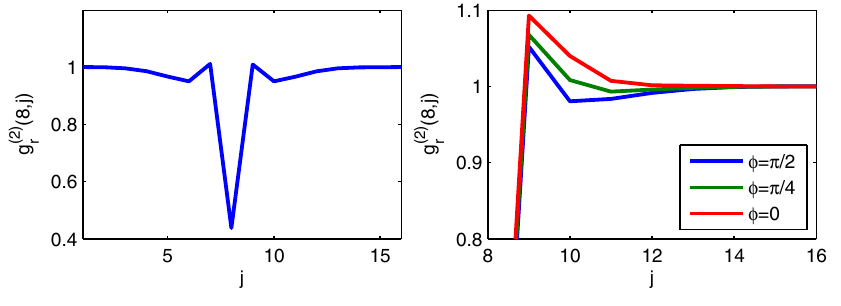}
\caption{$g_{8,j}^{(2)}$ for $\Delta=0$ and $U > \Omega > \gamma$ in a driven-dissipative Bose-Hubbard model. Right plot: Dependence of the correlations on the relative phase between adjacent driving fields. Reproduced from \cite{Hartmann10}.}
\label{fig:crystallisation}
\end{figure}

Given the expected relations to the Lieb-Liniger model, an interesting question is, whether a driven-dissipative Bose-Hubbard model exhibits similar density-density correlations, in particular whether a feature similar to Friedel oscillations \cite{Friedel:2007wt} can be expected.
Spatially resolved density-density correlations in the form of a $g^{(2)}$-function,
\begin{equation}
g_{j,l}^{(2)} = \frac{\langle a_j^\dag a_l^\dag a_l a_j \rangle}{\langle a_j^\dag a_j \rangle \langle a_l^\dag a_l \rangle}
\end{equation}
where $j$ and $l$ label lattice sites, have been investigated by Hartmann \cite{Hartmann10} via a numerical integration of equation (\ref{eq:master}) with Matrix Product Operators \cite{Schollwock:2011ys,Hartmann:2009jc}. A spatial modulation similar to Friedel oscillations was indeed found for a non-vanishing relative phase between the driving fields of adjacent lattice sites, see figure \ref{fig:crystallisation}. As this phase off-set generates a particle flux in the array, the strong anti-bunching on-site $g_{j,j}^{(2)} \ll 1$, slight bunching for neighbouring sites, $g_{j,j+1}^{(2)} > 1$, and anti bunching for further separated sites, $g_{j,l}^{(2)} < 1$ for $|j-l| > 1$, indicates that a flux of particle dimers extended over neighbouring sites flows through the lattice.
A similar signature was later also found for the driven-dissipative Jaynes-Cummings-Hubbard model by Grujic et al.  \cite{Grujic2012}.

\begin{figure}
\centering
\includegraphics[width=0.7\columnwidth]{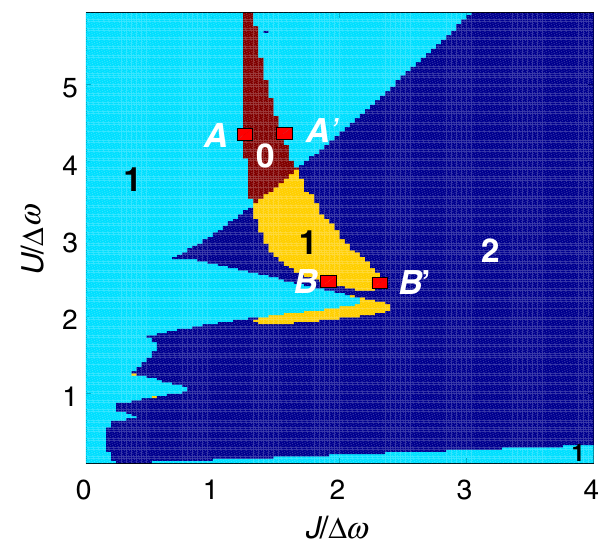}
\caption{Number of mean-field solutions for $\Omega/\Delta = 0.4$, $\gamma/\Delta = 0.2$ and $\Delta > 0$. Note that the notation $\Delta\omega = \Delta$ is used in the axes labels. Reproduced from \cite{Le-Boite:2013px}.}
\label{fig:ddBHbistable}
\end{figure}

For a higher dimensional lattice, a mean-field approach based on the exact single site solution by Drummond et al. \cite{Drummond:1980dp} was used by Le Boit\'e et al. \cite{Le-Boite:2013px,Le-Boite:2014ek} to predict mono - and bistable phases, which emerge as a consequence of the nonlinear nature of the mean-field equations, see figure \ref{fig:ddBHbistable}. 
The related dynamical hysteresis was then investigated in \cite{Casteels:2016kb}. Yet other calculation techniques rather show a first order phase transition in the regions where mean-field predicts a bistable phase \cite{Weimer:2015la,Maghrebi:2016la}.

In the limit of very large on-site interactions, the driven-dissipative Bose-Hubbard model maps to a driven-dissipative XY spin-1/2 model, the phase diagram of which has been investigated with a site-decoupled mean-field approximation \cite{Wilson16}
Here, stationary state phases with canted antiferromagnetic order and limit cycle phases with persistent oscillatory dynamics together with bistabilities of these two phases, have been found. 

\begin{figure}
\centering
\includegraphics[width=0.8\columnwidth]{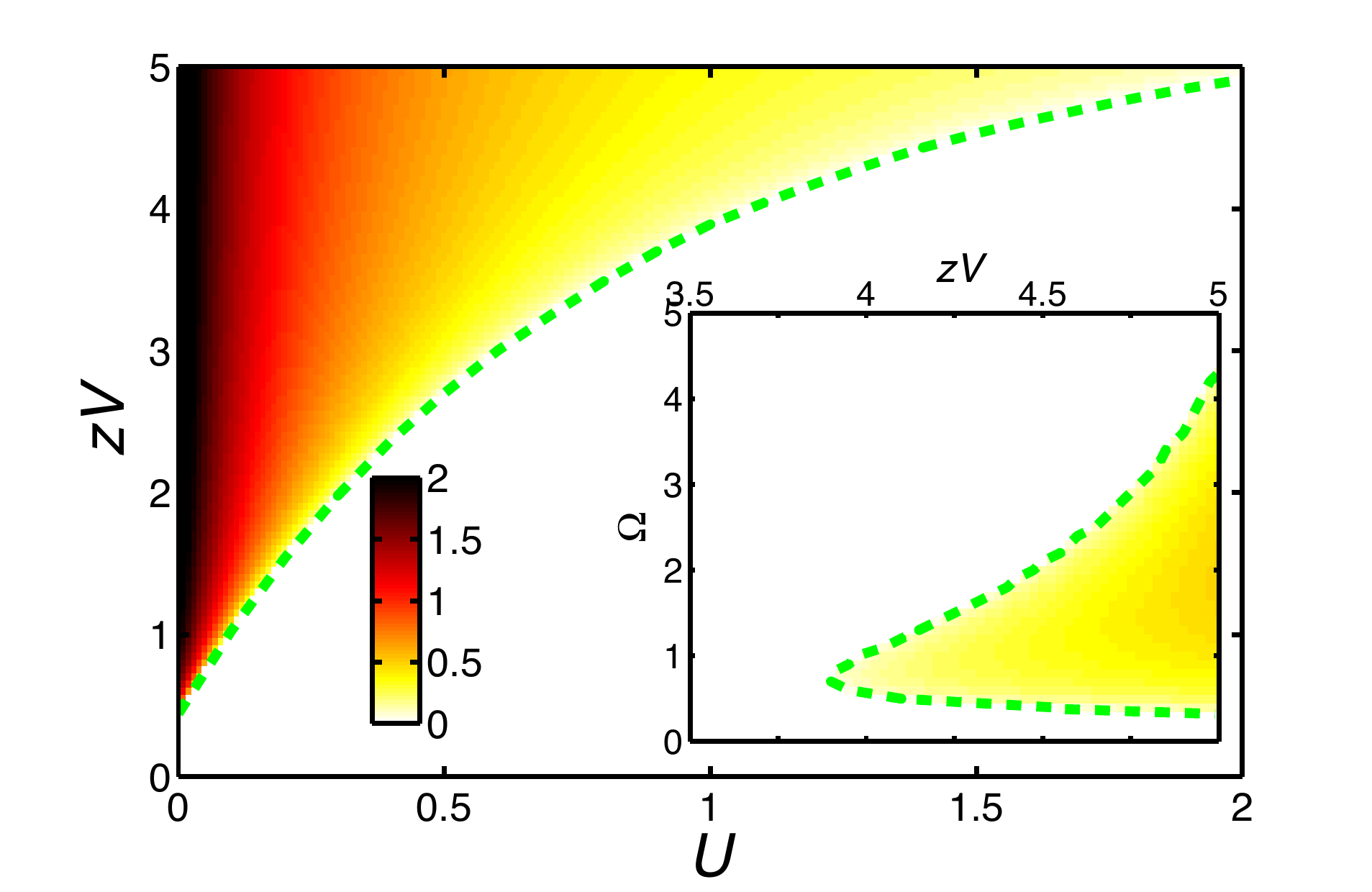}
  \caption{Order parameter $\Delta n$ of a mean-field calculation for a driven-dissipative Bose-Hubbard model
    with cross-Kerr interactions as described in equation (\ref{eq:crossKerr}) in the $U-V$ plane 
    at zero hopping. If the cross-Kerr term exceeds a critical threshold $V_c$, 
    the steady state is characterised by a staggered order in which $\Delta n \ne 0$. 
    Here we fixed $\Omega=0.75$ and $\Delta_{r} = 0$, for which $z V_c \approx 0.44$ at $U=0$, while 
    $zV_c \approx 5.73$ in the hard-core limit ($U \to \infty$).
    In the inset we show $\Delta n$ as a function of $\Omega$ and $V$ at a fixed value of $U = 1$.
    Here and in the next figure the colour code signals the intensity of the order parameter, 
    while dashed green lines are guides to the eye to locate the phase boundaries. Reproduced from \cite{Jin:2013qp}.}
\label{fig:BH-dens-wave}
\end{figure}

Moreover, driven-dissipative phases of the Bose-Hubbard model with cross-Kerr interactions as discussed in section \ref{sec:nonlocalints} were calculated with a mean-field technique for higher \cite{Jin:2014ax} and with Matrix Product Operators for one dimension \cite{Jin:2013qp}. For sufficiently strong cross-Kerr interactions and low enough excitation tunnelling, the model exhibits a density-wave ordering with $g^{(2)}(j,j+1) <  g^{(2)}(j,j)$, see figure \ref{fig:BH-dens-wave}.

In the studies discussed so far, the excitation dissipation has been assumed to be caused by the electromagnetic vacuum throughout.
As has been investigated in \cite{Quijandria:2015cs}, the scenario changes substantially if squeezed dissipation is considered, which does not obey a $U(1)$ symmetry.
We now turn to discuss driven-dissipative scenarios for the Jaynes-Cummings-Hubbard model.

\subsubsection{Driven-dissipative Jaynes-Cummings-Hubbard model} 
\begin{figure}
\centering
\includegraphics[width=\columnwidth]{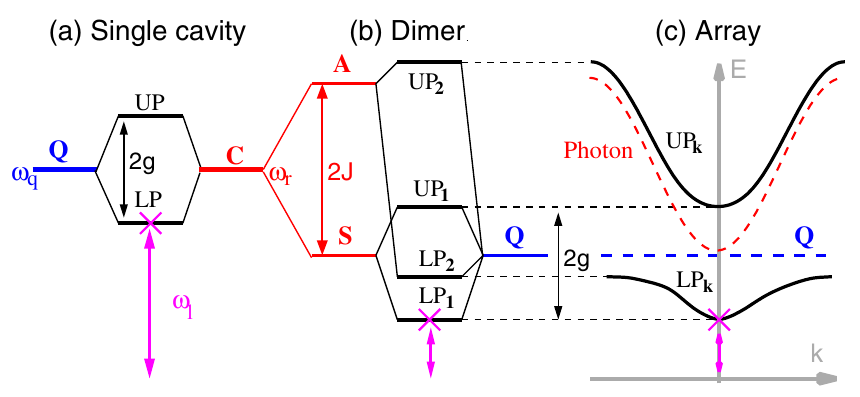}
\caption{Energy levels for the singe-excitation manifold of the Jaynes-Cummings-Hubbard model for a single site (a), a dimer (b) and a large array (c). The qubit transition frequency (Q) is chosen resonant with the symmetric two-cavity state (S) of the dimer and the bottom of the photon band of the array. Reproduced from \cite{Nissen}.}
\label{fig:JCspectra}
\end{figure}

$\,$
The coherence and fluorescence properties of a coherently pumped driven-dissipative Jaynes-Cummings-Hubbard model were explored by Nissen at al. \cite{Nissen}. For short arrays, the photon blockade regime was found to persist even up to large tunnelling rates, whereas there is a transition to a coherent regime for larger arrays as the tunnelling strength is increased. This size dependence is due to the fact that spectrally dense excitation bands only form in the limit of large system sizes whereas for a small lattice, say a dimer, the tunnelling causes a splitting of the spectral lines of single site spectra that leads to a collective spectrum which still remains anharmonic \cite{Nissen}, see figure \ref{fig:JCspectra}. 

A comparative study of the features of driven-dissipative Bose-Hubbard and Jaynes-Cummings-Hubbard models was conducted by Grujic et al. \cite{Grujic2012} and found quantitative differences for the experimentally accessible observables of both models for realistic regimes of interactions even when the corresponding nonlinearities are of similar strength.

Further interesting effects appear for arrays where not every resonator couples to a superconducting qubit but nonlinear interactions only occur in regularly spaced lattice sites. This periodic arrangement leads to photonic flat bands, see also section \ref{sec:artificial-gauge}, where the role of interactions is enhanced and polaritons can become incompressible \cite{Biondi:2015qm,Casteels:2016ye}. Experimentally, similar flat bands have been generated in coupled micro-pillars, each containing a cavity formed by two distributed Bragg reflectors \cite{Baboux:2016ys}, where condensation of polaritons in the flat band was observed. Such micro-pillars have also been coupled in a honey-comb lattice where they exhibit edge states similar to graphene structures \cite{Ozawa:2015wf}, c.f. section \ref{sec:artificial-gauge}. Whereas we have so far discussed scenarios with coherent driving fields, cavity arrays with incoherent input fields have been considered as well.

\subsubsection{Regimes of incoherent pumping}
In the above discussed driven-dissipative systems of interacting photons, the driving fields were always assumed to be of a coherent nature. Therefore any coherence between lattices sites that is found in the system needs to be attributed at least in part to the coherent inputs. To explore whether coherence can spontaneously develop for non-equilibrium photonic systems, incoherent input fields or pumping of higher excited states and subsequent decay \cite{Marcos2012} should be considered. A study for an incoherently pumped Jaynes-Cummings-Hubbard model found that the scaling of the correlation length with the hopping rate $J_{JC}$ changes as the hopping rate becomes larger than the light-matter coupling $g$ \cite{Ruiz-Rivas:2014qf}. See also the remarks in Sec. \ref{sec:microw-chem-pot} about engineering effective chemical potentials.
For further work on non-equilibrium photon condensation, see \cite{Klaers,Sieberer13,Altman:2015ef}, as well as the reviews \cite{Carusotto13,Sieberer2015} and references therein.

After reviewing approaches that considered a uniform drive intensity for all lattice sites, we now turn to discuss scenarios where the input can be more intense at one end of a chain. This setup naturally leads to the analysis of transport properties.

\subsubsection{Transport studies}
Transport of photons in a waveguide that are scattered at a localised emitter has been theoretically explored via scattering theory \cite{Shen2005,Liao:2010tk,Neumeier:2013bq}, see also extensions to multiple emitters \cite{Lee:2015ix},  and numerically using the Density Matrix Renormalization Group (DMRG) \cite{Longo2010}. Experimentally such scattering effects have been explored with Rydberg atoms \cite{Saffman:2010fq,Low:2012fk,Hofmann:2013uq,Chang:2014zt,Roy16} and one \cite{Astafiev2010,Hoi:2011ul} or two \cite{Loo:2013lq} superconducting qubits coupled to an open coplanar waveguide resonator. The recent developments in this direction of research are summarised in the review by Roy et al. \cite{Roy16}.
In our discussion in relation to quantum simulation we here therefore focus on transport studies of the driven-dissipative regime of a quantum many-body system on a one-dimensional chain or two-dimensional band. Here, a particle flux can either be generated by pumping locally at one end of the chain (the band) or by implementing a phase off-set between the driving fields at adjacent resonator in the direction of transport \cite{Hartmann10}.

A phase transition in the sense that two eigenvalues of the Liouvillian approach zero has been found for a spin chain with incoherent pump at one and losses at the opposite end by Prosen and Pi\v{z}orn \cite{Prosen:2008fk}. 
Hafezi et al. studied the propagation of few photon pulses in the polaritonic Lieb-Lininger model described in section \ref{sec:optical-continuous} by decomposing their wave-function in zero-, singe- and two-photon components \cite{Hafezi:2011xy,Hafezi:2012wq} and found that for an input that drives a single- or two-photon transition, the output will contain anti-bunched or bunched photons.
A different scenario with a coherent input at one end of the chain and uniform dissipation in all lattice sites has been studied by Biella et al. \cite{Biella:2015kk}, where the transport was found to be strongly influenced by the many-body resonances related to extended eigen-states of the chain. These transport properties can be interpreted as a generalisation of photon blockade, c.f. section \ref{sec:pbh}, to extended one-dimensional systems. More recently, Mertz et al. \cite{Mertz16} explored two scenarios, a source-drain setup with coherent drive at one end and enhanced dissipation at the opposite end of the chain, as well as the regime where relative phases between coherent inputs at neighbouring lattice sites generate a current in the presence of uniform dissipation. Employing a Gutzwiller mean-field approximation, the study considered two dimensional lattices with periodic boundary conditions in the direction perpendicular to the direction of transport. In addition to the dependencies of the current on the many-body spectrum it found that transport can be inhibited by strong dissipation at the ``drain''-end by the quantum Zeno effect. 

A relative phase between the coherent input fields at different locations can already lead to interesting effects when considered for the two outer resonators of a three-site model, which we turn to discuss now.     

\subsubsection{Josephson interferomenter}
The interplay of coherent tunnelling and on-site repulsion has been theoretically explored in three-cavity setups where the two outer cavities where driven by coherent fields the relative phase of which was varied. The central cavity in turn contained a Kerr nonlinearity, c.f. equation (\ref{eq:bosehubbard}). The first study coined the setup ``Quantum Optical Josephson Interferometer'' \cite{Gerace2009} and found that the destructive interference in the central cavity due to opposite phases of the driving fields is retained even for large nonlinearity. As photons enter the central cavity via tunnelling processes from the outer cavities, the transition from coherent to anti-bunched photon statistics in this cavity depends on the tunnelling rate. Similar behaviour was found for a case where two waveguides with a continuous spectrum replaced the two outer cavities. A later work explored the setup in a superconducting circuit context \cite{Jirschik:2014la}, c.f. section \ref{sec:microw-superc}, and extended the study to higher excitation numbers where a different dependence of the coherent to anti-bunching transition on the tunnelling rate was found.

Due to the large dimension of their Hilbert spaces and the eventually long time scales on which stationary states are reached, modelling many-body systems of interacting photons is a demanding challenge. In the next section we thus review some existing powerful methods together with recent efforts to extend their application ranges and develop new ones.

\subsection{Calculation Techniques}
\label{sec:calculation}
The modelling of quantum many-body systems is a formidable challenge since the dimension of their Hilbert space grows exponentially in the number of constituents. This scaling renders exact descriptions, including exact numerical approaches infeasible, even for moderate system sizes. For dissipative quantum many-body systems the computational effort is even more dramatic as mixed quantum states need to be considered. Exceptions to this intractability are quantum systems that do not explore their entire Hilbert space, where numerical optimisation approaches such as DMRG \cite{Schollwock:2011ys} become efficient. For calculating equilibrium phase diagrams DMRG has thus been used in a number of works considering many-body systems of interacting photons \cite{Rossini2007,Rossini:2012dq,Rossini:2008cr}. In turn for calculating dynamics, Matrix Product State representations of DMRG \cite{Orus:2014ul} in form of the Time-Evolving Block Decimation (TEBD) \cite{Vidal:2003yq,Vidal:2004vn,Zwolak:2004kx} have frequently been applied \cite{Hartmann:2008wo,Hartmann10,Grujic2012,Jin:2013qp,Peropadre:2013ul,Schetakis:2013bh,Quijandria:2014qf,Biondi:2015qm}.

Alternatively to approximations with states of limited entanglement, one may aim for only obtaining the information of interest about the quantum state of 
the entire system and try to find accurate and efficient approximations for the sought quantities.
Mean-field approaches \cite{Kadanoff:2009xy} can be understood as representatives of this strategy as they only predict properties of 
a single constituent of the many-body system \cite{Sachdev1999,Fisher89}. Such approaches, which calculate local quantities but ignore all correlations between subsystems, have been applied in both, the calculation of equilibrium states \cite{Greentree2006} as well as dynamics \cite{Tomadin2010a} and stationary states of driven-dissipative systems \cite{Nissen,Le-Boite:2013px}. In some cases, e.g. some scenarios of strongly interacting Rydberg gases as discussed in section \ref{sec:Rydberg}, only states with low excitation numbers contribute so that full numerics in a strongly truncated Hilbert space provides a good approximation \cite{Low:2012fk}. 

Yet, as DMRG approaches are limited to one-dimensional systems and mean-field techniques are only expected to become accurate in very high lattice dimensions, which often do not correspond to the physical realisations, there is a need for further efficient methods for accurately calculating stationary states of quantum many-body systems. As a consequence a substantial amount of research has recently been dedicated to the development of such methods. 

Keldysh path integral methods have been used to explore long-range properties \cite{Tauber:2014uq,Sieberer2015} and dynamical mean-field theory has been generalised to nonequilibrium scenarios \cite{Aoki:2014fy}. For solving Lindbald type master equations for their stationary states, del Valle et al. have expanded the resulting equations for correlators in powers of the inter-site coupling \cite{Valle:2013rz}. Li et al. have developed a perturbation theory for the Lindbladian including a resummation technique for the perturbations \cite{Li:2014uq,Li:2016fk}. Degenfeld-Schonburg et al. have generalised open quantum system techniques to take dynamical environments into account so that they can describe the interacting constituents of a many-body system in a consistent Mori projector theory (c-MoP) \cite{Degenfeld-Schonburg:2014bx,Degenfeld-Schonburg:2015tx}. 

For directly finding the stationary states of a master equation without doing a time integration, variational approaches have been developed. These include a variational expansion around product states \cite{Weimer:2015la} and variational Matrix Product Operator approaches \cite{Cui:2015pb,Mascarenhas:2015fy}. Alternative approximations that focus on a restricted part of the Hilbert space in a similar way as Matrix Product State representations by keeping only the dominant eigenvalues of reduced density matrices but can be applied in two-dimensional lattices  have been introduced by Finazzi et al. as a ``corner-space renormalization method'' \cite{Finazzi:2015fi}. Moreover, a dynamical polaron ansatz has been introduced for treating very strong light-matter couplings \cite{Diaz-Camacho15,Kurcz:2015lq}.

For experimental investigations of the predicted phase diagrams and phenomena, a crucial question is whether their experimental signatures are accessible in measurements. We therefore discuss some of these signatures in the next section.

\subsection{Experimental signatures}
\label{sec:phase-diag-signatures}
To clarify whether the phase diagrams and transitions discussed here can eventually be observed in experiments, it is important to determine their signatures in measurable observables. In this context it is natural to consider the photons emitted from the individual resonators. These output fluxes are related to intra-resonator quantities via input-output relations \cite{Wallsbook}. In this way the number of polaritons in each resonator,
$n_l = \langle p_l^{\dagger} p_l \rangle$ for site $l$, and the number fluctuations,
\begin{equation}
F_l = \langle (p_l^{\dagger} p_l )^2 \rangle - \langle p_l^{\dagger} p_l \rangle^2
\end{equation}
can be measured. These quantities provide information about the localisation and delocalisation of the polaritons, e.g. in a Mott insulator to superfluid transition.

The visibility of interference fringes of photons emitted from the resonators \cite{Rossini2007,Huo:2008qq} moreover provides information about coherences between resonators that may have built up in condensation, thus signalling a superfluid phase, c.f. \cite{BDZ07}. The visibility 
\begin{equation}
{\cal V} = \frac{\textrm{max}(S) - \textrm{min}(S)}{\textrm{max}(S) + \textrm{min}(S)} 
\end{equation}
of the interference pattern can be expressed in terms of the polariton number distribution in momentum space,
\begin{equation} \label{eq:momentum-dist}
S(k) = \frac{1}{L} \sum_{j,l = 1}^L e^{2 \pi i (j - l)/L} \Bra p_j^{\dagger} p_l \Ket,
\end{equation}
where $L$ is the number of sites in a one dimensional array.
In the superfluid regime, ${\cal V}$ approaches unity whereas it is very small for a Mott insulator. Remarkably, equation (\ref{eq:momentum-dist}) also allows to bound the entanglement in a many-body system from below without further assumptions \cite{Cramer:2011hx,Cramer:2013ts}.

Moreover, coincidence counting measurements of the photons emitted from one or multiple resonators \cite{Wallsbook} allow to reconstruct $g^{(2)}$-functions of the polaritons in the resonator for both, coincidences from one resonator leading to $g^{(2)}(j,j)$ and coincidences from separate resonators leading to $g^{(2)}(j,l)$ with $j \neq l$. These quantities would reveal the density wave ordering predicted for driven-dissipative Bose-Hubbard models with phase off-set in the driving fields \cite{Hartmann10} or with cross-Kerr interactions \cite{Jin:2013qp}. Higher order correlation-functions \cite{Valle:2012ev} could then be measured to reveal further properties of the investigated quantum many-body states.

In the next section we turn to review a recent development that has been largely triggered by the ample possibilities for engineering the band structure for photons in many devices. This is the realisation of artificial gauge fields for the quantum simulation of many-body systems under their influence.

\section{Artificial gauge fields}
\label{sec:artificial-gauge}
Charged particles moving in magnetic fields are an important paradigm in quantum mechanics and give rise to intriguing phenomena including the celebrated quantum Hall effect \cite{Yennie:1987lq,Stormer1999}. The increasing understanding of the geometric foundations of the quantum Hall effect also led to the discovery of topological insulators and topological superconductors which show exotic properties routed in the topological structure of their electron bands \cite{Qi:2011gf}.

The first approaches to exploring photons that are subject to an artificial gauge field were put forward by Haldane and Raghu \cite{Haldane2008,Raghu2008}, who considered a hexagonal array of dielectric rods leading to two dimensional photonic bands that show an appreciable Faraday effect leading to a breaking of time-reversal symmetry. The photonic bands can then be characterised by their Chern number \cite{Thouless:1982qc}, which is a topological invariant. Moreover, when two ``materials'' with different Chern numbers are joined, a chiral state emerges, that propagates along the interface in one unique direction only. These so called edge states are robust with respect to scattering at impurities provided the associated interaction energy is smaller than the gap to neighbouring bands. For photons these properties allow to engineer one-way waveguides which are free of backscattering. 

For microwave photons, an approach to engineer a time reversal symmetry breaking in networks of superconducting circuits has been introduced by Koch et al. \cite{Koch2010}, see also \cite{Nunnenkamp2011}. In this scheme, a circulator element formed by a superconducting ring intersected by three Josephson junctions that is threaded by a flux bias couples three coplanar waveguide resonators.

The absence of backscattering at impurities for edge modes has been considered by Hafezi et al. \cite{Hafezi:2011la} for engineering robust optical delay lines. In their approach, toroidal micro-cavities coupled by loops of tapered optical fibre with different path length for different directions of propagation are considered to emulate the effect of an artificial gauge field similar to a perpendicular magnetic field for the photon dynamics. An alternative approach to generating effective gauge fields via a dynamical modulation of the photon tunnelling rate at difference between the oscillation frequencies of the two adjacent resonators has been presented by Fang et al. \cite{Fang:2012bc}. 
The emerging gauge fields do in both concepts not need to be spatially uniform and can thus be employed to guide photon propagation and implement waveguides \cite{Lin:2014eb}.

Experimentally, the absence of backscattering in chiral edge modes of photons has first been investigated in the microwave regime in photonic crystals of a square lattice geometry \cite{Wang:2009kq}. Profiles of edge modes have then been imaged in silicon photonics devices \cite{HafeziM.:2013xy}. More recently, topological invariants in such systems have been measured via the shift of the spectrum as a response to a quantum of flux inserted at the edge \cite{Mittal:2016qf}. This technique allows to access the winding number which is directly related to the Chern number via the bulk-boundary correspondence. An alternative approach to engineering an artificial gauge field has been explored with a continuous drive on one site of a Bose-Hubbard dimer \cite{Rodriguez2016}.
For further details on the physics of artificial gauge fields for propagating photons, we refer the interested reader to the review by Hafezi \cite{HAFEZI:2013th}.

Whereas experimental progress has so far been mostly made with samples where photon-photon interactions via optical nonlinearities can be neglected, theory research has also addressed the intriguing regime where strong artificial gauge fields and strong effective photon-photon interactions coexist. Most notably this leads to regimes for fractional quantum Hall physics. A first proposal for generating a fractional quantum Hall regime in coupled photonic resonators was put forward by Cho et al. \cite{Cho2008}. The approach considers optical cavities doped with single atoms with a lambda shaped level structure featuring two metastable states. By driving two photon transitions with external lasers with nonuniform relative phases, an artificial gauge field is engineered in this strongly nonlinear system. This driving pattern leads to a bosonic version of a fractional quantum Hall regime, with even number of magnetic fluxes per excitation.

In circuit-QED systems in turn, three-body interactions in the presence of artificial gauge fields have been explored, which lead to Pfaffian states \cite{Hafezi:2014pb}. The artificial gauge field is in circuit-QED architectures implemented via externally modulated coupling SQUIDs as discussed in section \ref{sec:nonlocalints}.
The adjacent building blocks have different transition frequencies and the coupling SQUID is driven by an external flux which oscillates at a frequency that equals the difference of the transition frequencies of its two neighbouring lattice sites. Relative phases between the drives at several coupling SQUIDs then encode an applied gauge field. If in turn, the photon tunnelling is dynamically modulated at the sum of the transition frequencies of the adjacent lattice sites, a Kitaev spin chain showing Majorana zero modes can emerge \cite{Bardyn12}.

The dynamical modulation of a coupling circuit for engineering an artificial gauge field has recently been demonstrated in an experiment by Roushan et al. \cite{Roushan16}. Here three superconducting transmon qubits were coupled in a ring via tunable couplers \cite{Chen:2014ao} that were modualted by oscillating magnetic fluxes to simulate the effect of a perpendicular gauge field. In this setup, that thus combines an artificial gauge field with local interactions provided by the qubit nonlinearities, single excitations and excitation pairs were circulated around the ring in a controlled way.

The emergence of fractional quantum Hall physics in driven-dissipative lattices of coupled photonic resonators has been investigated by Umucalilar et al.  \cite{Umucalilar12} by showing that the stationary states of such driven dissipative systems can, when projected on a specific excitation number, approximate Laughlin wave-functions with even number of magnetic fluxes per excitation. The overlap of the stationary states of such cavity lattices with a Laughlin state was analysed with an efficient approximation for low excitation numbers in \cite{Hafezi:2013tt}.
The potential of engineered dissipation for stabilising topological states by coupling the cavity lattice to two-level systems with fast dissipation, has been explored by Kapit et al. \cite{Kapit:2014fy}. Moreover a quantum simulator for topological order with superconducting circuits has been proposed in \cite{Sameti16}

The theory research on models with artificial gauge fields has recently also been pushed further to explore the generation of non-Abelian \cite{Mezzacapo:2015yb} and dynamical gauge fields in lattice gauge filed theories \cite{Marcos:2013cr,Marcos:2014et}. For the latter, the gauge fields are not set by an external current or voltage source by formed by dynamical degrees of freedom of the network.

\section{Summary and Outlook}
\label{sec:summary}
After optical nonlinearities at the single photon level, i.e. effective interactions between individual photons, have been realised in single micro-cavities, coupling several optical or microwave resonators to form a network has become a new research goal. In parallel avenues to generate effective interactions between individual photons in extended one-dimensional volumes have been considered. Whereas the initial work in the research field was mostly of theoretical nature, technological advances in recent years have now matured the experimental platforms to such an extent that an increasing number of experimental investigations are to be seen in the coming years. This development has two exciting perspectives.

From a scientific perspective, quantum many-body systems of interacting photons can be expected to exhibit a wealth of new quantum many-body phenomena as they naturally operate under driven, non-equilibrium conditions which are different to the equilibrium scenarios usually explored in quantum many-body physics.
From a technology perspective, photons are the most suitable carrier for transmitting information over long distances as they are largely immune to environmental perturbations. As optical nonlinearities in conventional media are weak it is therefore of great importance to conceive means of making individual photons interact with each other at multiple nodes of a network to make them suitable for information processing. 

\ack
The author thanks Fernando Brand\~{a}o, Peter Degenfeld-Schonburg, Elena del Valle, Martin Leib, Martin Kiffner, Lukas Neumeier, Alessandro Ridolfo, Rosario Fazio, Salvatore Savasta and particularly Martin Plenio for collaborations in projects that are part of the research covered in this review. Moreover he is grateful for discussions during the KITP programme ``Many-body physics with light'' that provided helpful input for this review and wants to thank Dimitris Angelakis, Cristiano Ciuti, Iacopo Carusotto, Sebastian Diehl, Mohammad Hafezi, Jonathan Keeling, Jens Koch, Lucas Lamata, Giovanna Morigi, Peter Rabl, Juan Jos\'{e} Garc\'{i}a Ripoll, Davide Rossini, Sebastian Schmidt, Enrique Solano, Hakan T\"ureci and Hendrik Weimer for discussions and comments.
This work was supported by the EPSRC under EP/N009428/1 and supported in part by the National Science Foundation under Grant No. NSF PHY11-25915

\section*{References}
\bibliographystyle{plain}
\bibliography{QuSim}

\end{document}